\documentclass[twocolumn]{aa} 
\usepackage{multirow}
\usepackage{amsmath}
\usepackage{subfigure}
\usepackage{epic,eepic}
\usepackage{graphicx}
\usepackage{txfonts}
\usepackage{hyperref}
\usepackage{gensymb}
\usepackage{textcomp, gensymb}
\usepackage{orcidlink}
\usepackage{appendix}
\usepackage{booktabs}
\usepackage{placeins}  
\usepackage{caption}

\begin{document}

\title{Ca~{\sc ii} IRT activity indices from {\it Gaia} spectroscopy: a baseline method for photospheric contribution estimates}

\titlerunning{Photospheric Contribution to Ca~{\sc ii} IRT emission}

\author{Song Wang\inst{1,2}\thanks{Corresponding author: songw@bao.ac.cn}
        \and
        Tanda Li\inst{2,3}
        \and
        Henggeng Han\inst{1}
        \and
        Xue Li\inst{4}
        \and
       Jifeng Liu\inst{4,5}
      }

\institute{National Astronomical Observatories, Chinese Academy of Sciences, Beijing 100101, China
\and
Institute for Frontiers in Astronomy and Astrophysics, Beijing Normal University, Beijing 102206, China
\and
School of Physics and Astronomy, Beijing Normal University, Beijing, 100875, China
\and
School of Astronomy and Space Sciences, University of Chinese Academy of Sciences, Beijing 100049, China
\and
New Cornerstone Science Laboratory, National Astronomical Observatories, Chinese Academy of Sciences, Beijing 100101, China
}

\date{Received 15 July 2026 / Accepted 12 November 2026}

\abstract{
The {\it Gaia} high-resolution spectroscopic observations using Radial Velocity Spectrometer offers a unique opportunity to study stellar Ca~{\sc ii} infrared triplet (IRT) activity. A key step in estimating activity indices is subtracting the photospheric contribution. The canonical template-based method often fails because, for many stars, the observed line cores are deeper than the model predictions, which prevents valid activity measurements.
In this work, we explore an alternative approach that constructs a baseline from less active stars as a function of effective temperature (and metallicity) to represent the photospheric contribution. 
Our results show that, for dwarf stars, compared with the canonical method, the baseline method not only yields far more objects with valid activity measurements, but also avoids many extremely low and unreliable values (e.g., log$R'_{\rm IRT} \lesssim -6$).
The Ca~{\sc ii} IRT activity indices derived from the baseline method correlate well with other activity proxies, confirming their reliability as activity indicators. Furthermore, they show a relation with stellar rotation, not only recovering the typical rotation--activity relation, but also revealing fine structures that may be associated with different regimes in gyrochronology.
Our findings suggest that the baseline method can serve as a robust alternative for deriving accurate activity estimates, and will be particularly useful for obtaining activity indices for tens of millions of stars with the upcoming {\it Gaia} Data Release 4.}

\keywords{stars: atmospheres -- stars: chromospheres -- stars: activity}

\maketitle

\section{Introduction}

Magnetic activity of late-type stars provides insight into magnetic energy release processes operating at different atmospheric layers, which are observable as different phenomena (e.g., star spots, chromospheric plages, ultraviolet emission, and X-ray emission) across the electromagnetic spectrum \citep[e.g.,][]{2008A&A...479..557F,2009A&ARv..17..251S,2011ApJ...743...48W,2013MNRAS.431.2063S}.
One of the best-known measures of magnetic activity is the Mount-Wilson S-index  ($S_{\rm MWO}$) and the derived $R^{'}_{\rm HK}$, calculated using the flux in the center of the Ca~{\sc ii} H and K lines.
The Ca~{\sc ii} infrared triplet (Ca~{\sc ii} IRT) lines at $\lambda\lambda$ 8498, 8542, 8662\,\AA\ share the upper level of the H and K transitions and closely resemble them in many respects, indicating that they can also serve as chromospheric activity indicators \citep{2000A&A...353..666C,2000A&AS..146..103M,2007A&A...466.1089B}.

{\it Gaia} mission performed high-resolution spectroscopic observations with the Radial Velocity Spectrometer (RVS), covering a wavelength range of 8450--8720\,\AA\ at a resolution of $R\approx11,500$, for 32.2 million stars brighter than $G\sim$15\,mag \citep{2023A&A...674A...6S}.
Considering that activity indices calculated from spectra are strongly affected by spectral resolution and that low-resolution spectra require further calibration \citep{2025ApJ...984....2H}, these high-resolution spectra provide a great opportunity to perform accurate Ca~{\sc ii} IRT activity measurements.
\citet{2023A&A...674A..30L} presented stellar activity indices derived from the Ca~{\sc ii} IRT lines for two million stars in the Milky Way, using both published and unpublished RVS spectra.

Generally, the calculation of Ca~{\sc ii} IRT activity indices involves two steps: integration of the flux in the cores of the Ca~{\sc ii} IRT lines and removal of the photospheric contribution. Previous studies have developed different methods to perform these two steps. For example, \citet{2017A&A...605A.113M} used an integration range of 1\,\AA\ around the core of each of the triplet lines, while \citet{2019A&A...623A..24F} and \citet{2023A&A...674A..30L} adopted a 0.5\,\AA\ and 1.5\,\AA\ interval, respectively. \citet{2007A&A...466.1089B} even used a variable integration range, from 0.1\,\AA\ in slow-rotating, non-active stars up to 6\,\AA\ in fast rotators and/or very active stars. 
For the photospheric contribution, previous studies have employed local thermodynamic equilibrium (LTE) \citep{2023A&A...674A..30L} or non-LTE (NLTE) \citep{2005A&A...430..669A,2007A&A...466.1089B,2017A&A...605A.113M,2024ApJS..272....6H} synthetic spectral models. 
The resulting quantity after subtracting the photospheric contribution can be called excess equivalent width ($EW$).
In addition to these two steps, \citet{2023A&A...674A..30L} further converted the excess $EW$ to $R'_{\rm IRT}$ to express the flux excess over the radiative equilibrium approximation in the Ca~{\sc ii} IRT cores in bolometric units.

However, previous methods for estimating the photospheric contribution have some disadvantages. 
The most significant issue is that, for many stars, the observed line cores are deeper than those predicted by the parameter-matched templates. \citet{2026ApJS..287...38Y} interpreted this discrepancy as arising primarily from the omission of chromospheric structure in templates, with a secondary contribution from NLTE effects. As a result, a large proportion of stars are prevented from yielding meaningful activity measurements.

In this study, we introduced a simple method to estimate the photospheric contribution (as well as the chromospheric emission unrelated to magnetic activity) using a baseline constructed from less active stars, which was also mentioned by \citet{1993ApJS...86..293D}. 
We adopted two approaches to build this baseline, with and without accounting for metallicity, since metallicity can influence the emission levels of metal lines.
The paper is organized as follows.
Section~\ref{data.sec} describes the dataset and the methods for estimating the Ca~{\sc ii} IRT activity indices. 
Section~\ref{result.sec} presents the results and discussion, covering the comparison of indices from different methods, the correlation between Ca~{\sc ii} IRT and other activity proxies, and the rotation--activity relation derived from Ca~{\sc ii} IRT activity.
Finally, a summary is given in Sect. \ref{sum.sec}.

\section{Dataset and method}
\label{data.sec}

{\it Gaia} Data Release 3 (DR3) has published high-resolution spectra (i.e., {\tt rvs\_mean\_spectrum}) for 999,645 stars. 
These mean spectra were generated from multiple individual observations, each of which was first normalized to the pseudo-continuum and shifted to the rest frame before combination.
We downloaded all these spectra from the {\it Gaia} archive data website\footnote{\url{https://cdn.gea.esac.esa.int/?prefix=Gaia/gdr3/Spectroscopy/rvs_mean_spectrum}}. By visual inspection, we found that the continua of some spectra are not well normalized. Therefore, we re-normalized all spectra before the subsequent analysis.
We adopted {\tt laspec.normalization.normalize\_spectrum\_spline} for normalization, which uses a cubic smoothing spline to model the continuum.

\subsection{Atmospheric parameters}

Atmospheric parameters are needed during the calculation of the activity indices. 
Gaia DR3 release presents two sets of stellar parameters: the General Stellar Parametrizer from spectroscopy (GSP-Spec) using high-resolution RVS spectra, and the GSP from photometry (GSP-Phot) using low-resolution BP/RP spectra.
\citet{2023A&A...674A..30L} used the parameters from GSP-Spec by default, since the activity index was derived from the same data; when they were not available, the ones from GSP-Phot were used. 

Here, we did not simply follow this approach. Instead, we first assessed which set of parameters is more reliable, with the Large Sky Area Multi-Object Fiber Spectroscopic Telescope (LAMOST) parameter catalog \citep{2025ApJS..279....5Z} as a reference. 
This catalog contains atmospheric parameters for 6.4 million stars released in LAMOST low-resolution DR9, derived with the DD-PAYNE method.
We cross-matched the LAMOST and {\it Gaia} DR3 catalogs and selected the common sources for which LAMOST, GSP-Spec, and GSP-Phot all provided valid values.
There are 97,381 common stars between LAMOST and GSP-Spec, and 91,004 common stars between LAMOST and GSP-Phot.

\begin{figure*}[htbp]
\begin{center}
\includegraphics[width=0.95\textwidth]{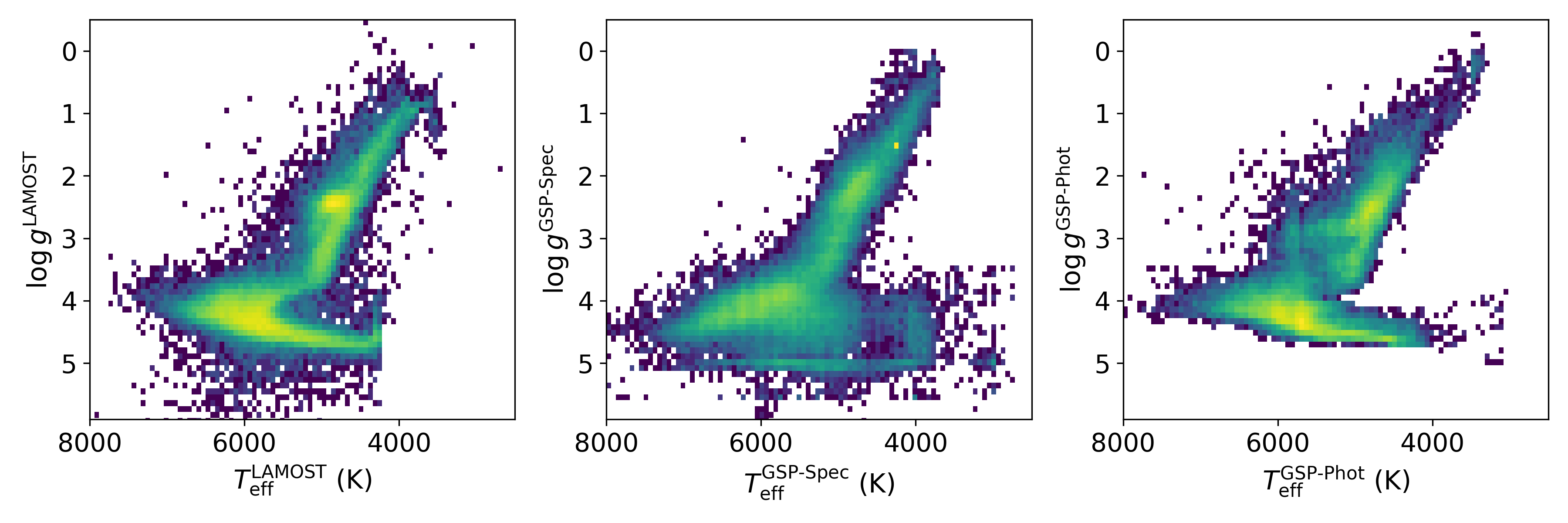}
\noindent
\captionof{figure}{Kiel diagram using different parameter sets: LAMOST (left panel), {\it Gaia} GSP-Spec (middle panel), and GSP-Phot (right panel).}
\label{tefflogg.fig}
\end{center}
\end{figure*}

\begin{figure*}[htbp]
\begin{center}
\includegraphics[width=0.95\textwidth]{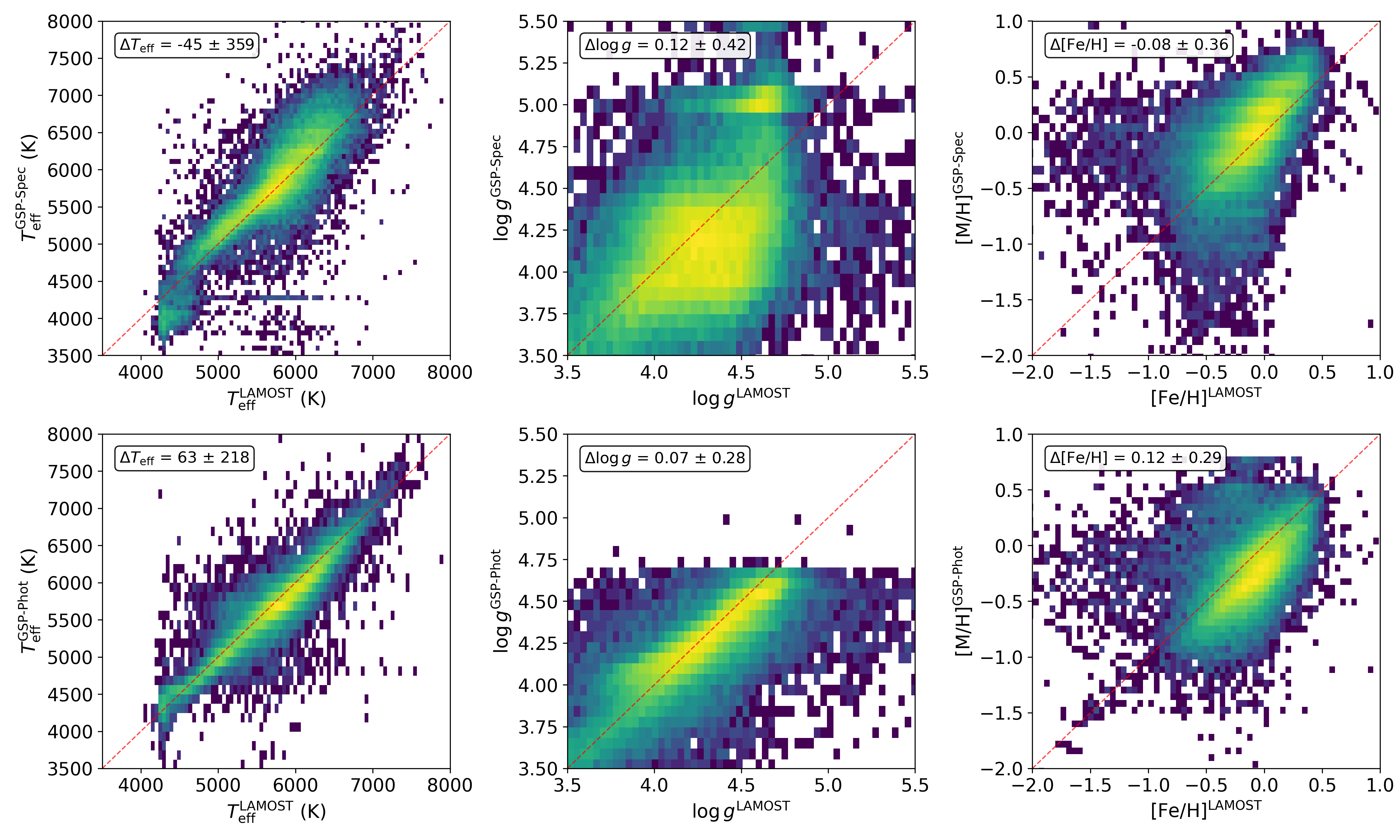}
\noindent
\captionof{figure}{Comparison of atmospheric parameters between LAMOST and GSP-Spec (top panels), and between LAMOST and GSP-Phot (bottom panels) for stars with log$g$ values (in all three datasets) larger than 3.5.}
\label{comlamost.fig}
\end{center}
\end{figure*}

\begin{figure}[t]
\begin{center}
\includegraphics[width=0.48\textwidth]{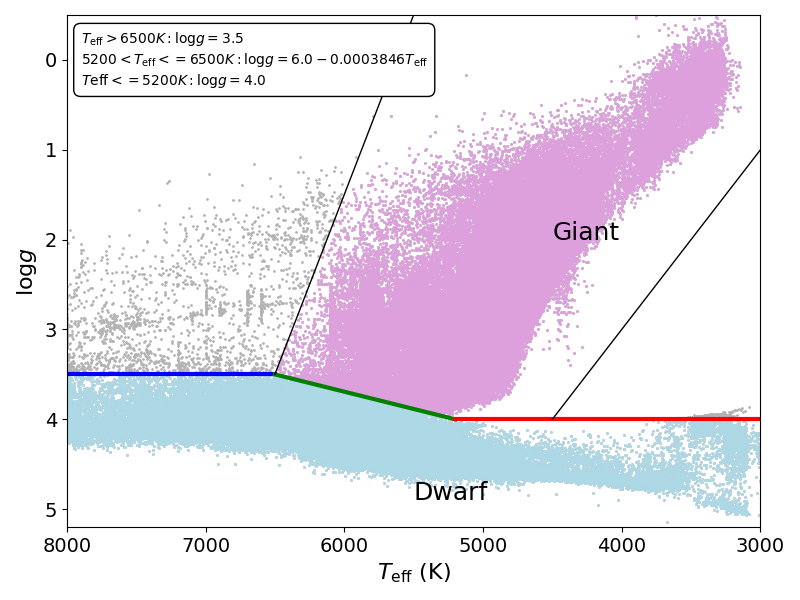}
\caption{Kiel diagram of our sample. Blue, green, and red lines mark the separation between dwarfs (blue points) and giants (plum points). The light gray points are excluded from the sample.
\label{logg.fig}}
\end{center}
\end{figure}

Figure \ref{tefflogg.fig} presents the $T_{\rm eff}$--log$g$ diagram (also called Kiel Diagram), constructed using the common stars. The LAMOST and GSP-Phot parameters define a tighter sequence than the GSP-Spec parameters, especially for main-sequence stars.
Figure \ref{comlamost.fig} compares the atmospheric parameters from LAMOST with those from GSP-Spec and GSP-Phot for stars with log$g$ values (in all three datasets) larger than 3.5 (e.g., main-sequence stars and some subgiants). 
The temperature difference between LAMOST and GSP-Spec is smaller than that between LAMOST and GSP-Phot, but with substantially larger scatter. 
The surface gravities from GSP-Phot are in good agreement with LAMOST, showing much better consistency than those from GSP-Spec. 
GSP uses [M/H] as its metallicity, while LAMOST uses [Fe/H]. We neglected this difference in the comparison.
The metallicity difference between LAMOST and GSP-Spec is smaller than that between LAMOST and GSP-Phot, again with larger scatter.
Therefore, we used a hybrid combination of parameters from GSP-Spec and GSP-Phot: the temperature and log$g$ values were taken from GSP-Phot, while the metallicity values were taken from GSP-Spec. 
Specifically, the parameters {\tt teff\_gspphot}, {\tt logg\_gspphot}, and {\tt mh\_gspspec} in {\it Gaia} DR3 catalog were used for stellar classification (i.e., dwarf or giant) and for baseline estimation.
A total of 437,782 stars (44\%) have valid measurements for all three parameters. Using these parameters, we classified the stars into dwarfs and giants, as shown in Fig. \ref{logg.fig}.

\subsection{Sample cleaning}

We removed binaries and other contaminants, such as galaxies and active galactic nuclei (AGNs), by cross-matching our sample with several existing catalogs.
First, we collected catalogs with binary classifications from light curve database, including Catalina \citep{2014ApJS..213....9D}, 
All-Sky Automated Survey for Supernovae \citep[e.g.,][]{2018MNRAS.477.3145J, 2023MNRAS.519.5271C}, 
General Catalog of Variable Stars \citep{2017ARep...61...80S}, 
Zwicky Transient Facility \citep{2020ApJS..249...18C}, 
Wide-field Infrared Survey Explorer \citep{2018ApJS..237...28C}, 
Transiting Exoplanet Survey Satellite \citep[TESS;][]{2022RNAAS...6...96H, 2022ApJS..258...16P, 2025ApJS..276...57G,2026ApJS..284...75B}, 
the Washington Double Star Catalog \citep{2001AJ....122.3466M},
the Ninth Catalog of Spectroscopic Binary Orbits \citep{2004A&A...424..727P},
and {\it Gaia} \citep{2023A&A...674A..14R}.
Second, we removed sources listed in the {\it Gaia} DR3 non-single source catalog \citep{2023A&A...674A..34G}. 
Third, stars with RUWE values (from {\it Gaia} DR3 catalog) larger than 1.4, which were possibly binaries, were removed.
Finally, we cross-matched the sample with SIMBAD \citep{2000A&AS..143....9W} and excluded contaminants including binaries, galaxies, and AGNs. 
A matching radius of 3\,arcsec was adopted for all cross-matching.

\begin{figure*}[ht!]
\begin{center}
\includegraphics[width=0.95\textwidth]{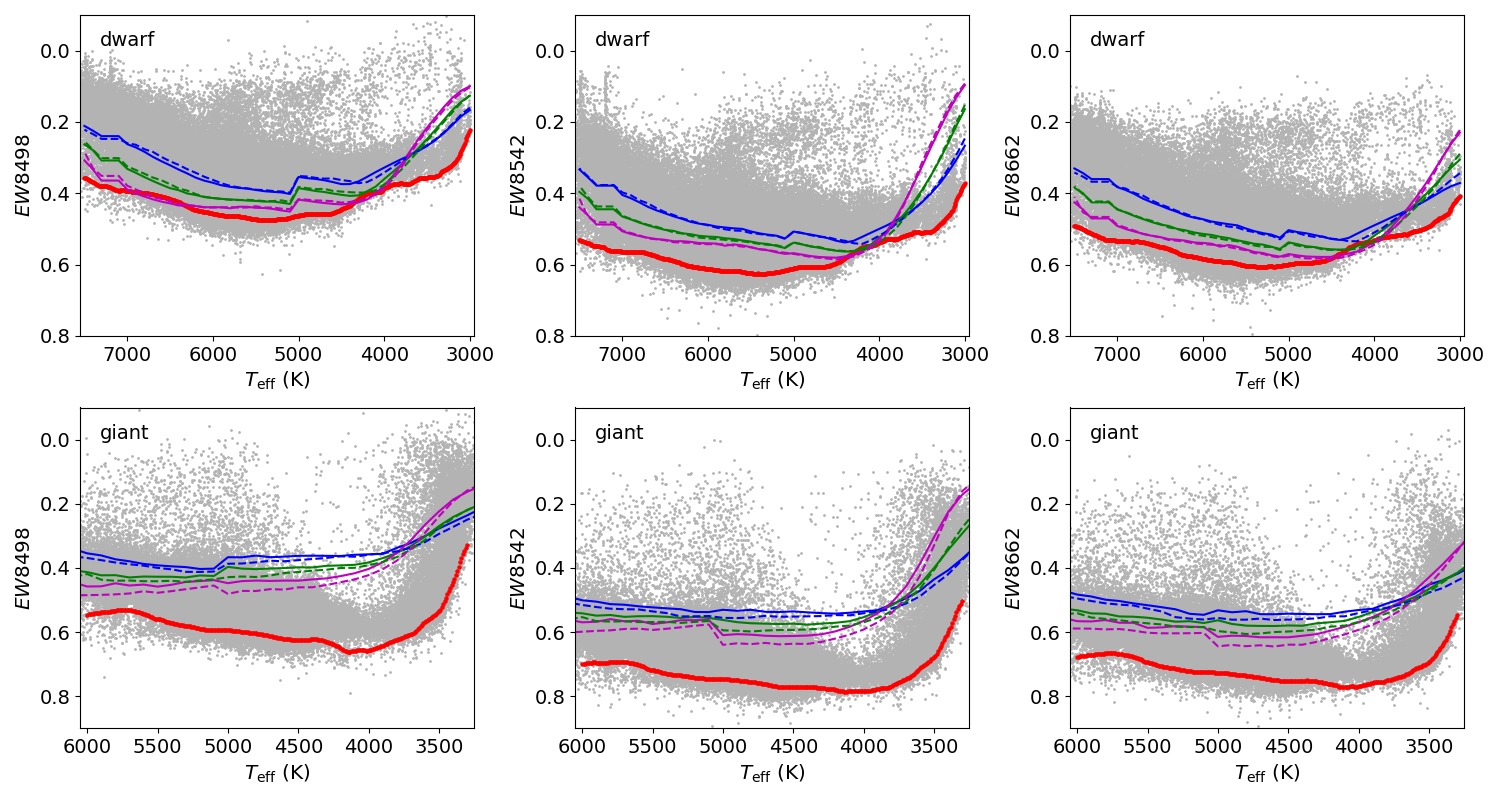}
\caption{Baseline construction (Method 2) for each Ca~{\sc ii} IRT line, for dwarfs (top panels) and giants (bottom panels), using less active stars as a function of effective temperature ($T_{\rm eff}$). 
$EW$8498, $EW$8542, and $EW$8662 represent $EW$ values of Ca~{\sc ii} $\lambda\lambda$ 8498, 8542, and 8662\,\AA\ lines, respectively.
Thick red lines represent the baselines. For comparison, blue, green, and purple lines represent the $EW$ values calculated from PHOENIX models with [Fe/H] values of $-$0.5, 0, and 0.5, respectively. For dwarfs, solid and dashed lines correspond to log$g =$ 4.5 and 4.0, respectively; for giants, solid and dashed lines correspond to log$g =$ 3.0 and 2.5, respectively.
\label{ew_base.fig}}
\end{center}
\end{figure*}

\begin{figure*}[ht!]
\begin{center}
\includegraphics[width=0.95\textwidth]{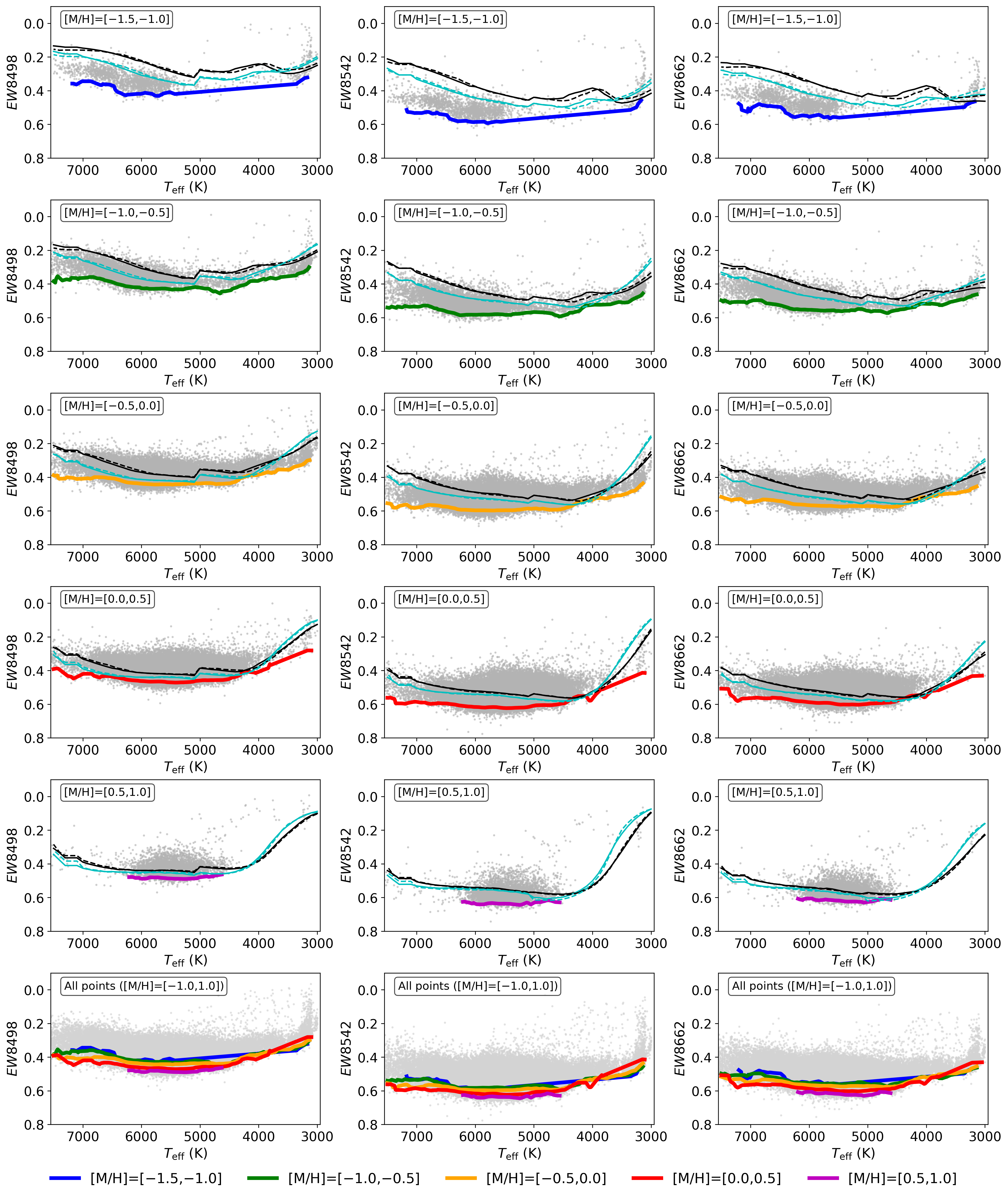}
\caption{Baseline construction (Method 3) for each Ca~{\sc ii} IRT line for dwarfs, using less active stars as a function of effective temperature ($T_{\rm eff}$) and metallicity ([M/H]). 
Thick lines denote the baselines. For comparison, black and cyan lines represent the $EW$ values calculated from PHOENIX models with [Fe/H] values corresponding to lower and upper boundaries of each metallicity bin, respectively. 
Solid and dashed lines correspond to log$g =$ 4.5 and 4.0, respectively.
\label{ew_base_dwarf_feh.fig}}
\end{center}
\end{figure*}

\subsection{Calculation of $R^{'}_{\lambda}$ and $R^{'}_{\rm IRT}$}

$R^{'}_{\rm IRT}$ is the normalized Ca~{\sc ii} IRT luminosity index after subtracting the photospheric radiation contribution. 
The activity indices of each Ca~{\sc ii} IRT line are denoted as $R^{'}_{\lambda 8498}$, $R^{'}_{\lambda 8542}$, and $R^{'}_{\lambda 8662}$, respectively.

First, we calculated the $EW$ of each Ca~{\sc ii} IRT line as follows,
\begin{equation}
EW_{\lambda}=\int 1- \frac{F_\lambda}{F_c} \, d\lambda.
\end{equation}
The integration range was set to 1\,\AA $\;$ around the line centers of the three spectral lines (i.e., 8500.35\,\AA, 8544.44\,\AA, and 8664.52\,\AA) following \citet{2017A&A...605A.113M} and \citet{2024ApJS..272....6H}.
Here, $F_c$ was taken as the median value of the continuum, defined over regions located 10--30\,\AA\ from the line center on both sides of each line.
Using $EW$ or simply integrating $F_\lambda / F_c$ in previous studies are equivalent, since the constant unity cancels out when the photospheric contribution is subtracted.

Second, we used three methods to estimate the photospheric contribution: 
(1) the canonical method using template matching \cite{2005A&A...430..669A}; 
(2) a baseline constructed from less active stars as a function of temperature  \citep{1993ApJS...86..293D};  
(3) a baseline constructed as a function of both temperature and metallicity. Unlike the canonical method, the baseline approach is less suitable for giant stars, because giants exhibit a wide range of log$g$ values, which can significantly affect the depth of the spectral lines. 
In this paper, we therefore applied the baseline methods only to dwarfs, while the template method was used for both dwarfs and giants.

In Method 1, we used the PHOENIX model \citep{2013A&A...553A...6H} to calculate the photospheric contribution ($EW_{\rm phot}$).
We built a model grid of $EW_{\rm phot}$ for $T_{\rm eff}$ between 2500 and 8000\,K in steps of 100\,K, log$g$ between 3 and 7 in steps of 0.5, [Fe/H] between $-$2.0 and 1.0 in steps of 0.5, and {\tt vbroad} between 0 and 50\,km s$^{-1}$ in steps of 5\,km s$^{-1}$.
For each star, we adopted the {\tt vbroad} parameter in the {\it Gaia} DR3 catalog as the observed projected rotational velocity.
For stars with observational parameters within the grid, we applied a four-dimensional interpolation to calculate $EW_{\rm phot}$.

In Methods 2 and 3, we built the baselines using the observed $EW$ values.
The baseline constructions for these two methods are shown in Figs. \ref{ew_base.fig} and \ref{ew_base_dwarf_feh.fig}, respectively.
We selected the top 5\% of sources with the highest $EW$ values (i.e., less active) and adopted their median as the baseline. Then, a grid of basal $EW$ values, corresponding to $EW_{\rm phot}$, was constructed as a function of temperature for Method 2 and of temperature plus metallicity for Method 3.
Due to the lack of sufficient stars in some grid cells, the baseline is only defined over a narrow temperature range. For example, in the metallicity bin of 0.5 $<$ [M/H] $<$ 1.0, baseline estimates are only available for stars with temperatures between $\sim$4500 and $\sim$6000\,K (Fig. \ref{ew_base_dwarf_feh.fig}).
For stars with observational parameters within the grids, we applied one-dimensional and two-dimensional interpolation to calculate $EW_{\rm phot}$ for Methods 2 and 3, respectively.

The photospheric contribution derived from template matching using PHOENIX models is notably higher than the baselines (Figs. \ref{ew_base.fig} and \ref{ew_base_dwarf_feh.fig}), indicating that many stars have observed line depths deeper than those predicted by the templates (i.e., the observed line fluxes are lower), supporting recent findings of \citet{2026ApJS..287...38Y}.

Third, we calculated the excess $EW$ related to chromospheric activity for each Ca~{\sc ii} IRT line following
\begin{equation}
EW_{\lambda}^{'}=EW_{\lambda}-EW_{{\lambda},\rm phot}.
\end{equation}
The activity index $R^{'}_{\lambda}$ was then calculated following 
\begin{equation}
R'_{\lambda} = \frac{L_{\lambda}}{L_{\mathrm{bol}}} = \chi \cdot EW_{\lambda}^{'},
\end{equation}
where $\chi$ was estimated as
\begin{equation}
\chi = \frac{f_{\lambda }}{f_{\mathrm{bol}}} = \frac{f_{\lambda }}{\sigma T_{\mathrm{eff}}^4}.
\end{equation}
The continuum flux $f_{\lambda }$ was estimated using the PHOENIX synthetic spectra. For each source, we selected the synthetic spectrum that best matches its effective temperature, surface gravity, and metallicity, performed a continuum fit, and adopted the continuum flux at each Ca~{\sc ii} IRT line core as $f_{\lambda }$.

Finally, we calculated $R^{'}_{\rm IRT}$ by averaging $R^{'}_{\lambda 8498}$, $R^{'}_{\lambda 8542}$, and $R^{'}_{\lambda 8662}$.
All $R^{'}_{\lambda}$ and $R^{'}_{\rm IRT}$ values derived from Methods 1, 2, and 3 are listed in Tables \ref{m1.tab}, \ref{m2.tab}, and \ref{m3.tab}, respectively.

\section{Results and discussion}
\label{result.sec}

\begin{figure*}[t]
\begin{center}
\includegraphics[width=0.98\textwidth]{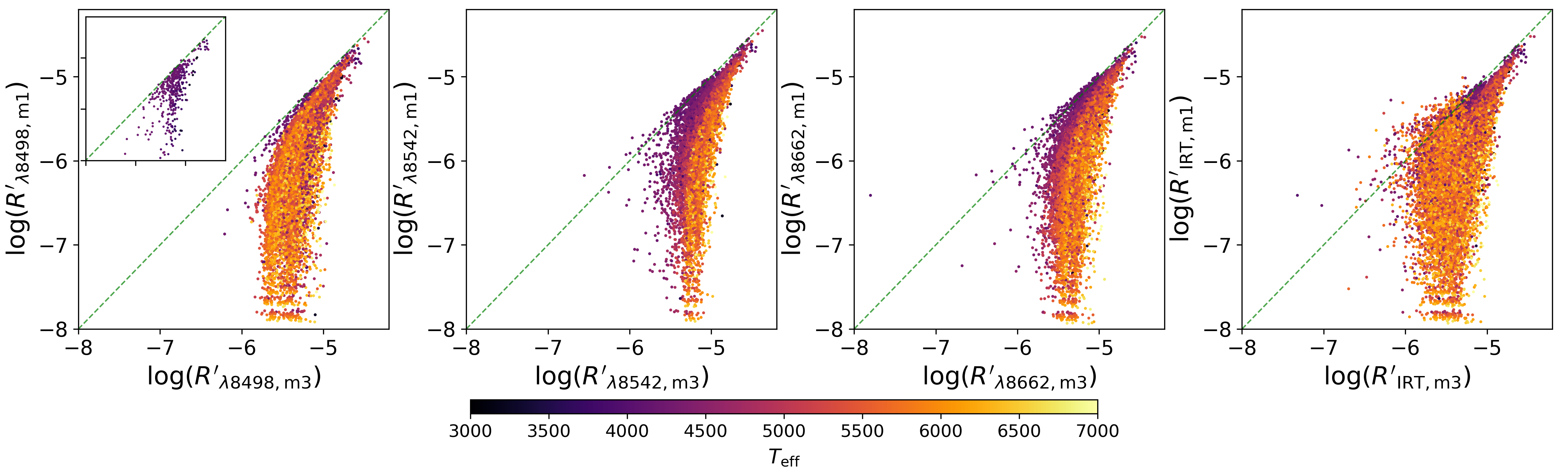}
\includegraphics[width=0.98\textwidth]{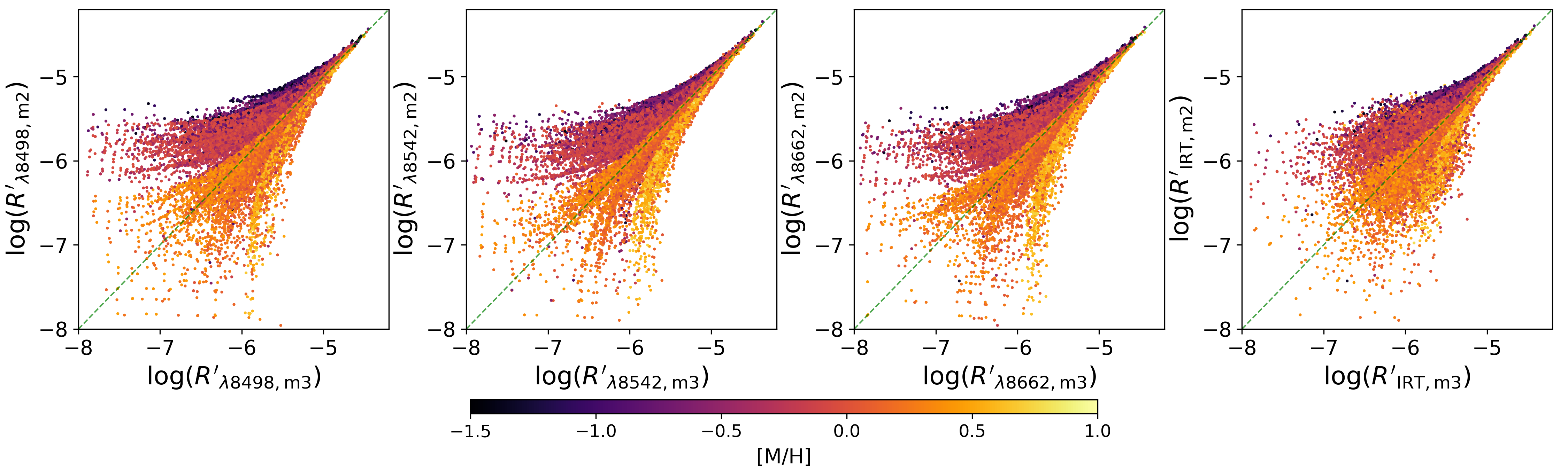}
\caption{Comparison of $R'_{\lambda}$ and $R'_{\rm IRT}$ values between Methods 1 and 3 (top panels) and between Methods 2 and 3 (bottom panels). The inset shows stars with temperatures lower than 4300\,K.
\label{rcom.fig}}
\end{center}
\end{figure*}

\subsection{Comparison of $R^{'}_{\rm IRT}$ among different methods}

Figure \ref{rcom.fig} (top panels) compares the $R'_{\lambda}$ and $R'_{\rm IRT}$ values derived using Methods 1 and 3. As discussed above, the $EW$ values derived from template matching are smaller than the observed $EW$ values for a significant fraction of stars. As a result, not only do many stars fail to yield valid $R'_{\lambda}$ and $R'_{\rm IRT}$ measurements, but the derived $EW'$ values are also much smaller than those obtained with the baseline methods, leading to systematically lower $R'_{\lambda}$ and $R'_{\rm IRT}$ values.

At temperatures around 4000\,K, the template-derived $EW$ values closely match the baseline-derived values, with increasing difference toward both lower and higher temperatures (Figs. \ref{ew_base.fig} and \ref{ew_base_dwarf_feh.fig}). Accordingly, for stars near 4000\,K, the $R'_{\lambda}$ and $R'_{\rm IRT}$ values from Methods 1 and 3 are most consistent.

Figure \ref{rcom.fig} (bottom panels) presents a comparison of $R'_{\lambda}$ and $R'_{\rm IRT}$ values derived from Methods 2 and 3. It is clear that Method 2 yields higher values for metal-poor stars and lower values for metal-rich stars. This is expected, as the intrinsically shallower Ca~{\sc ii} IRT absorption lines in metal-poor stars would lead to an apparent enhancement in the measured activity \citep{1991ApJ...375..722S}. 
In Method 2, the baseline was constructed without considering metallicity; therefore, the metallicity effect remains imprinted in the activity values. 
In contrast, Method 3 incorporates both temperature and metallicity in the baseline construction, effectively weakening the metallicity effect. 
Such metallicity dependence has also been noted in many studies of Ca~{\sc ii} H\&K activity \citep[e.g.,][]{2018A&A...619A..73L, 2025ApJ...983L..31C, 2025ApJ...984....2H}.

With the same initial sample, Method 1 returns only 32,374 valid activity measurements (Table~\ref{m1.tab}), including 2,526 giants and 29,848 dwarfs. By comparison, Methods 2 and 3 return 253,586 and 178,081 valid measurements for dwarfs (Tables \ref{m2.tab} and \ref{m3.tab}), respectively. Thus, our baseline method enables the construction of a much larger sample of activity measurements for statistical studies.

\begin{figure*}[ht!]
\begin{center}
\includegraphics[width=0.98\textwidth]{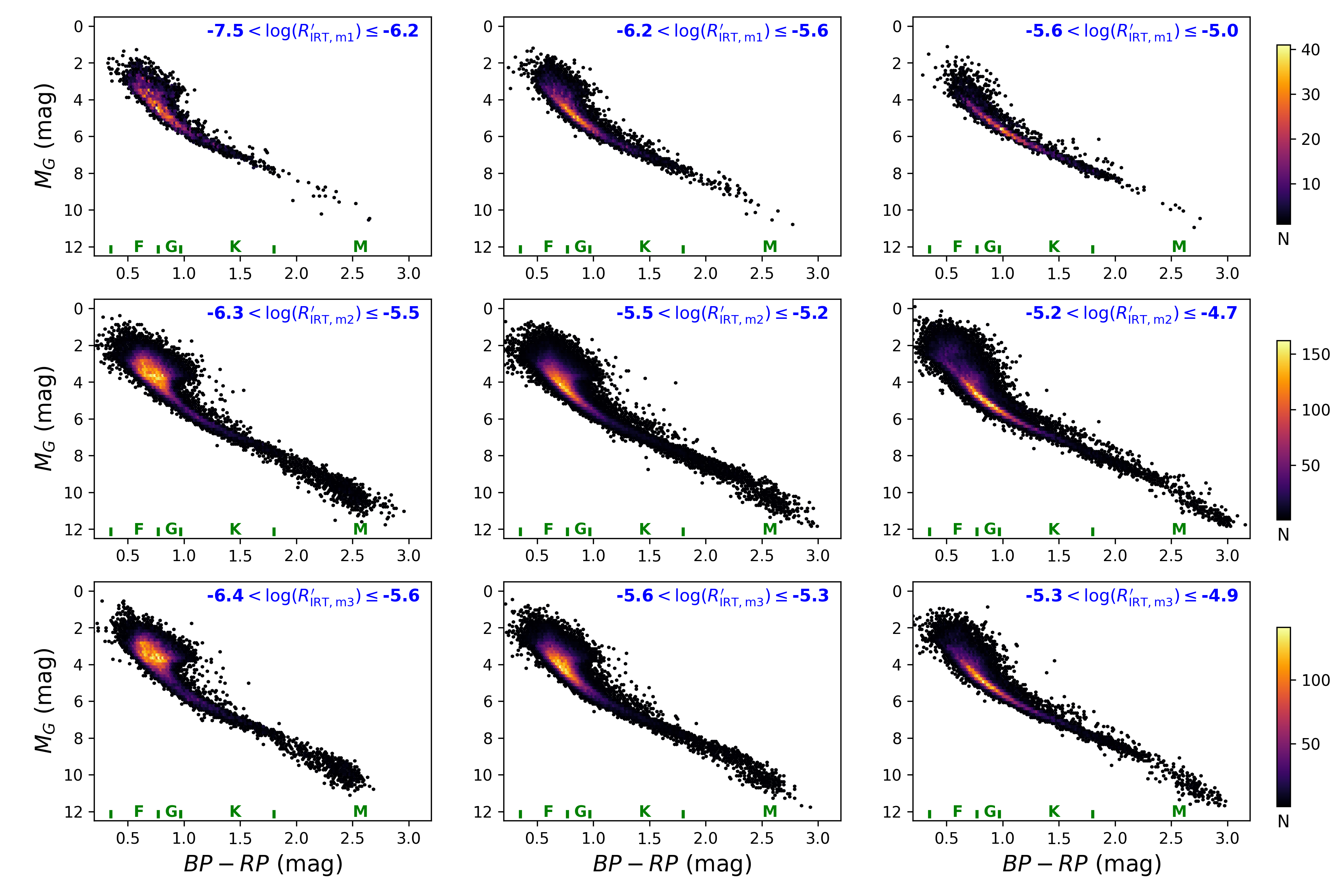}
\caption{Hertzsprung-Russell diagram of the sample stars, with colors representing the number of stars in each bin. 
Results for Methods 1, 2, and 3 are shown in the top, middle, and bottom panels, respectively.
The left, middle, and right panels show stars with $R'_{\rm IRT}$ values in the 1\%--25\%, 25\%--75\%, and 75\%--99\% percentiles, respectively.
\label{hr.fig}}
\end{center}
\end{figure*}

Figure \ref{hr.fig} shows the Hertzsprung-Russell diagram for our sample dwarfs, which is dominated by F and G stars, with only a few M dwarfs. 
The Method 1 results (top panels) contain about 46\% F, 46\% G, 6\% K, and 0.06\% M stars. In comparison,  the proportions for Method 2 (middle panels) are about 61\%, 29\%, 4\%, and 0.4\%, while those for Method 3 (bottom panels) are around 62\%, 31\%, 4\%, and 0.3\%.
A group of possible unresolved binaries (or pre-main-sequence stars) lies above the main sequence. As activity increases (from left to right panels), our sample shifts to lower-mass stars.

\subsection{Comparison of $R^{'}_{\rm IRT}$ with other proxies}
\label{other.sec}

The strength of magnetic activity can be traced by many proxies, including spots, flares, chromospheric emission (e.g, Ca H\&K, H$\alpha$, and ultraviolet), and X-ray and radio emission.
Generally, these proxies are correlated with each other \citep[e.g.,][]{2002ARA&A..40..217G,2004A&ARv..12...71G,2008ApJ...687.1264M,2013MNRAS.431.2063S,2020ApJ...902..114W}.

\begin{figure*}[!htbp]
\begin{center}
\includegraphics[width=0.95\textwidth]{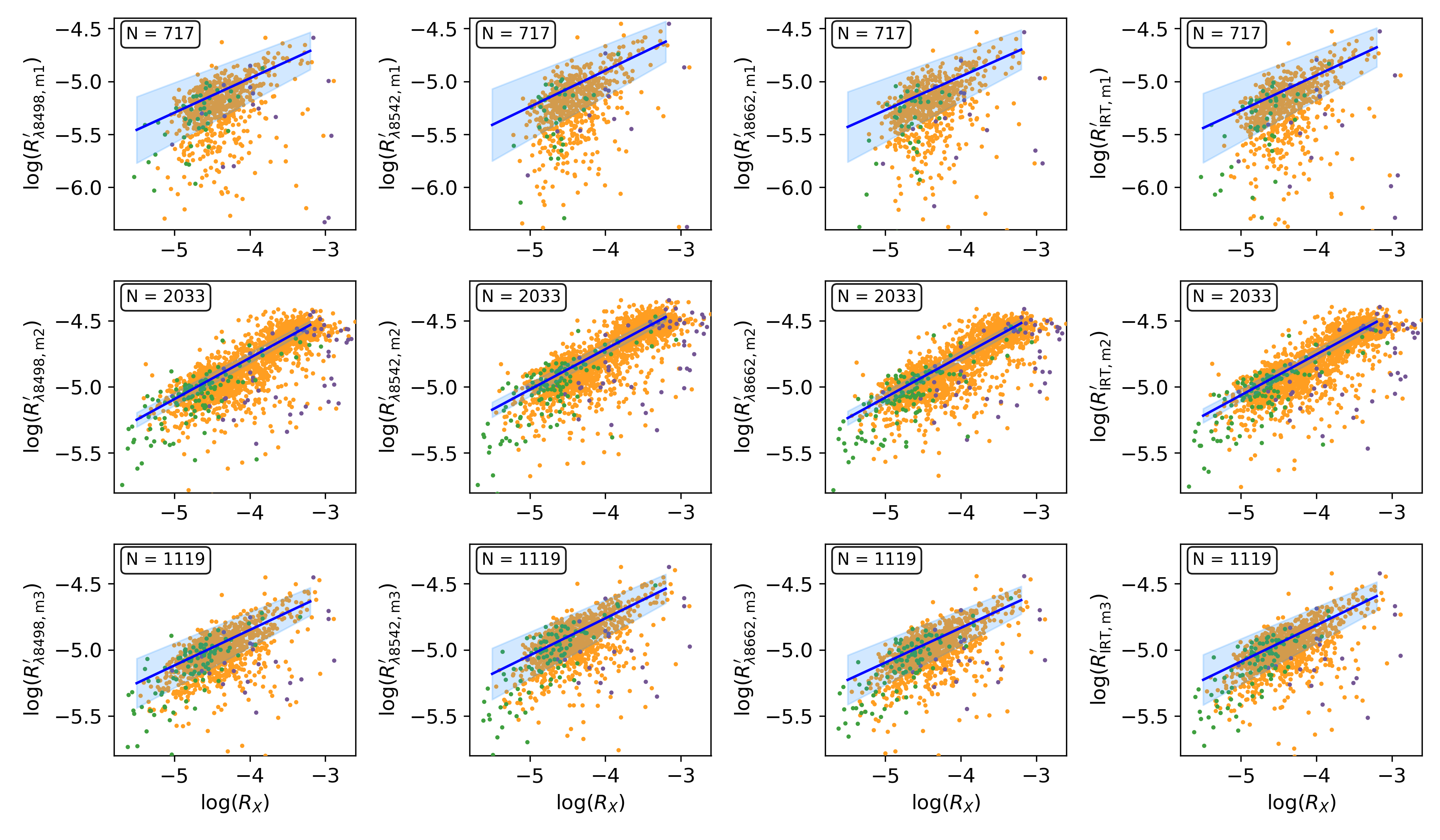}
\caption{Comparison between $R_X$ from the literature and $R'_{\lambda}$ and $R'_{\rm IRT}$ derived from Method 1 (top panels), Method 2 (middle panels), and Method 3 (bottom panels). X-ray data were taken from {\it ROSAT} (green points), {\it Chandra} and {\it XMM-Newton} (purple), and {\it eROSITA} (orange) observations.
\label{rcomx.fig}}
\end{center}
\end{figure*}

We compiled a catalog of stars with $R_X$ ($=L_X/L_{\rm bol}$) estimates derived from the literature, including {\it ROSAT} \citep{2024ApJS..273....8H}, {\it Chandra} and {\it XMM-Newton} \citep{2026ApJS..286...71L}, and {\it eROSITA} \citep{2025A&A...697A.230F}. 
Figure \ref{rcomx.fig} shows the comparison of X-ray activity proxy $R_X$ with the $R'_{\lambda}$ and $R'_{IRT}$ values derived from different methods.
The Ca~{\sc ii} IRT activity correlates positively with X-ray activity, with the results from Methods 2 and 3 showing tighter relations than those from Method 1.
We carried out linear fits for each comparison, and the resulting best-fit parameters are given in Table \ref{fit.tab}.
As can be seen, the uncertainties of fitted parameters for Methods 2 and 3 are much smaller than those for Method 1.

The number of common stars between the X-ray sample and Method 2 is larger than that with Methods 1 and 3, because the latter two require both {\tt teff\_gspphot} and {\tt mh\_gspspec} values, which are only available for 44\% of the full sample. Moreover, Method 1 fails for many stars because the PHOENIX templates have shallower lines than the observed spectra.
We also noticed that the stars uniquely covered by Method 2 (i.e., not by Methods 1 or 3) tend to show higher Ca~{\sc ii} IRT and X-ray activities (e.g., $R'_{\rm IRT} \gtrsim -$4.7 and $R_X \gtrsim -$3.5). 
Further inspection of these active objects reveals that they all have effective temperatures lower than 6000\,K.

We constructed a catalog of stars with $R'_{\rm HK}$ measurements from the literature, based mainly on \citet{2018A&A...616A.108B}.
We noticed that some stars of \citet{2018A&A...616A.108B} have different names or multiple $R'_{\rm HK}$ values. For each source, we collected all available values and derived the median. Although \citet{2018A&A...616A.108B} already included sources from \citet{2010ApJ...725..875I}, we found that some stars from the latter were missing. We therefore supplemented the $R'_{\rm HK}$ measurements with sources from \citet{2010ApJ...725..875I}, as well as \citet{2024ApJS..274...35I} and \citet{2025AJ....170..212S}. 

\begin{figure*}[!htbp]
\begin{center}
\includegraphics[width=0.95\textwidth]{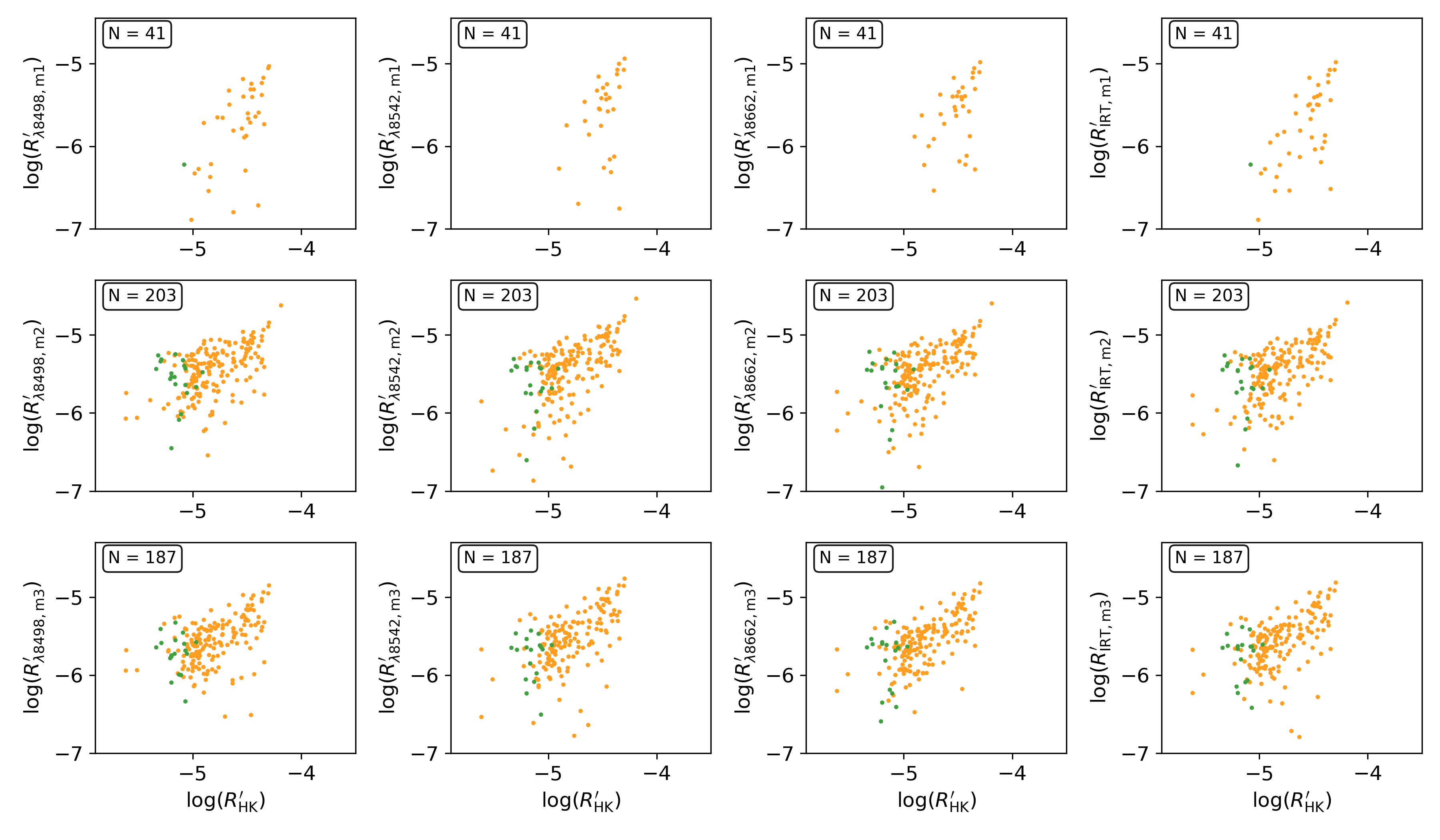}
\caption{Comparison between $R'_{\rm HK}$ from the literature and $R'_{\lambda}$ and $R'_{\rm IRT}$ derived from Method 1 (top panels), Method 2 (middle panels), and Method 3 (bottom panels). Orange and green points are from \citet{2018A&A...616A.108B} and other studies, respectively.
\label{rcomhk.fig}}
\end{center}
\end{figure*}

Figure \ref{rcomhk.fig} shows the comparison of the literature $R'_{\rm HK}$ values with the $R'_{\lambda}$ and $R'_{IRT}$ values derived from different methods.
Overall, they are positively correlated. However, due to the small sample size and the scatter, we did not attempt to fit relations for these quantities.

\subsection{Comparison of $R^{'}_{\rm IRT}$ with {\it Gaia} activity indices}

\citet{2023A&A...674A..30L} presented a catalog of Ca~{\sc ii} IRT activity indices (denoted as {\tt activityindex\_espcs} or $\alpha$) for two million sources.
They integrated the flux in the cores of each Ca~{\sc ii} IRT line over a 1.5\,\AA\ window and removed the photospheric contribution using template matching based on stellar effective temperature, surface gravity, and metallicity.
They also provided a conversion from $\alpha$ into $R^{'}_{\rm IRT}$ that depends on both temperature and metallicity (their Table 1 and Eq. 7), applicable for stars with metallicities between $-1.0$ to 1.0 and effective temperatures between 3000 to 7000\,K.

\begin{figure*}[ht!]
\begin{center}
\includegraphics[width=0.95\textwidth]{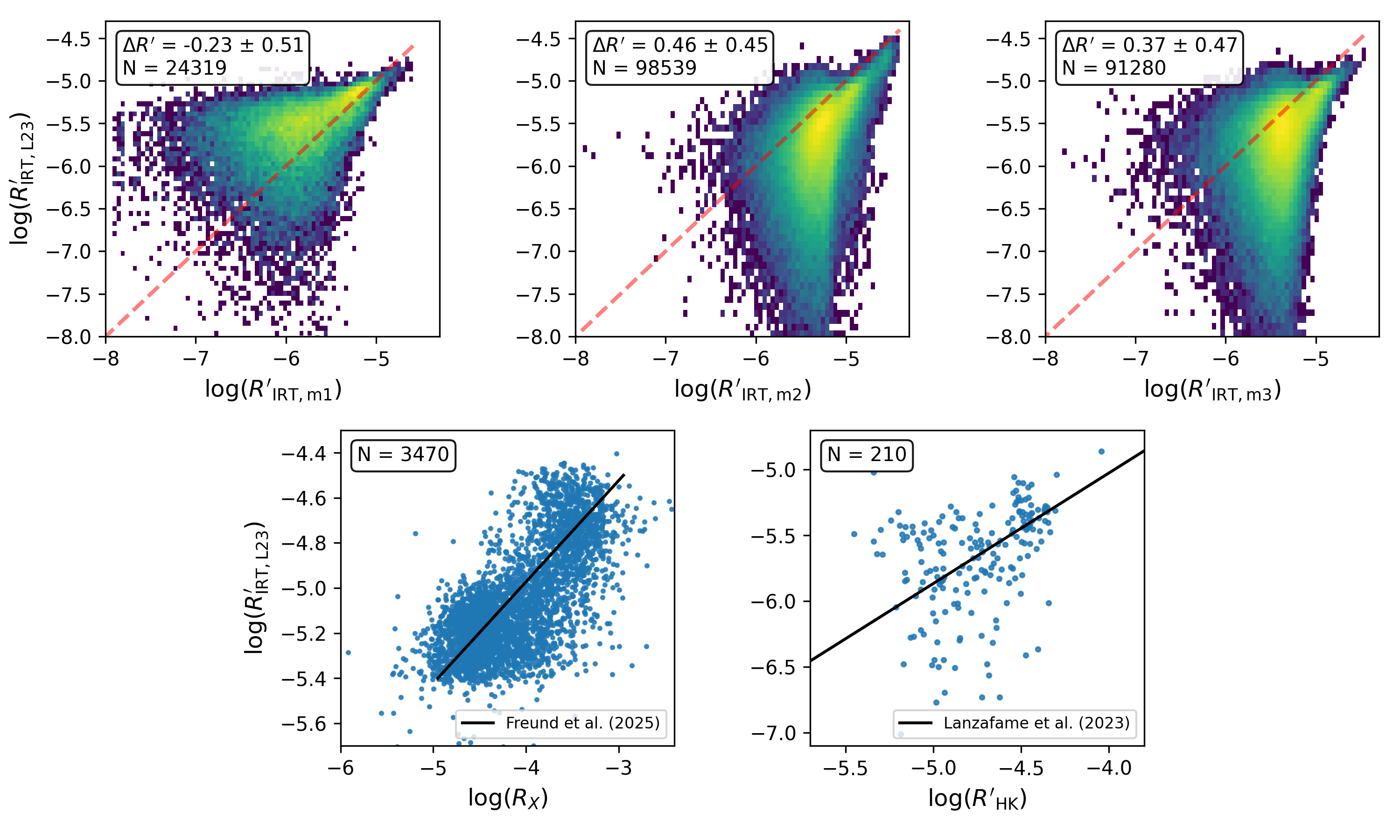}
\caption{Top panels: Comparison of $R'_{\rm IRT, L23}$ from the {\it Gaia} DR3 catalog with our $R'_{\rm IRT}$ values (derived from Methods 1, 2, and 3). Colors indicate the number of stars. Bottom panels:  Comparison of $R'_{\rm IRT, L23}$ from the {\it Gaia} DR3 catalog with the literature $R_X$ and $R'_{\rm HK}$ values.
The equation from \citet{2025A&A...697A.230F} is ${\rm log}(R_X)=2.224{\rm log}(R'_{\rm IRT})+7.060$; the equation from \citet{2023A&A...674A..30L} is ${\rm log}(R'_{\rm IRT})=0.84102786{\rm log}(R'_{\rm HK})-1.660991$.
\label{rcomg.fig}}
\end{center}
\end{figure*}

We cross-matched our initial sample of one million sources with that catalog, and calculated $R'_{\rm IRT, L23}$ values (from $\alpha$) for about 0.18 million sources. 
Figure \ref{rcomg.fig} (top panels) compares $R'_{\rm IRT, L23}$ values with our $R'_{\rm IRT}$ values derived from Methods 1, 2, and 3.
Since the method of \citet{2023A&A...674A..30L} is similar to our Method 1, the two agree well in the comparison.
In contrast, the comparisons between $R'_{\rm IRT, L23}$ and our Methods 2 and 3 show that $R'_{\rm IRT, L23}$ is systematically lower, consistent with the behavior seen when comparing Method 1 with Methods 2 and 3 (Fig. \ref{rcom.fig}).

Different studies can yield different chromospheric activity measurements, since the data and procedures involved can differ considerably, such as spectral resolution, normalization method, integration width, pseudocontinuum definition, and photospheric correction.
However, both \citet{2023A&A...674A..30L} and our Method 1 include many stars with indices between $-6.5$ to $-$8.0. About 33\% of the Method 1 results (Table~\ref{m1.tab}) fall in this range, compared with only 2\% and 3\% for Methods 2 and 3 (Tables~\ref{m2.tab} and~\ref{m3.tab}), respectively.
We suggested that the discrepancy is driven primarily by different methods used to correct for the photospheric contribution.
The very low $R'_{\rm IRT, L23}$ values from \citet{2023A&A...674A..30L} were also noted by \citet{2025A&A...697A.230F}, who suggested that stars with log$R^{'}_{\rm IRT, L23}$ $<$ $-5.4$ may have unreliable indices.

We also cross-matched the $R'_{\rm IRT, L23}$ values of the 0.18 million sources with the literature $R_X$ and $R'_{\rm HK}$ data described in Section \ref{other.sec}.
The $R'_{\rm IRT, L23}$ vs. $R_X$ diagram is similar to that shown by \citet{2025A&A...697A.230F}, and a simple linear function does not appear to fit the data well.
For example, for $R'_{\rm IRT, L23} \lesssim -4.9$, $R_X$ increases with $R'_{\rm IRT, L23}$, while for $R'_{\rm IRT, L23} \gtrsim -4.9$, $R_X$ remains nearly constant at a high level.
Due to the small sample size, the $R'_{\rm IRT, L23}$ vs. $R'_{\rm HK}$ comparison shows notable scatter. In fact, \citet{2023A&A...674A..30L} also showed a large scatter between these two quantities. In their Fig. 7, many stars have $R'_{\rm IRT, L23}$ values lower than $-6$, and exhibit poor correlation with $R'_{\rm HK}$.

\begin{figure*}[ht!]
\begin{center}
\includegraphics[width=0.95\textwidth]{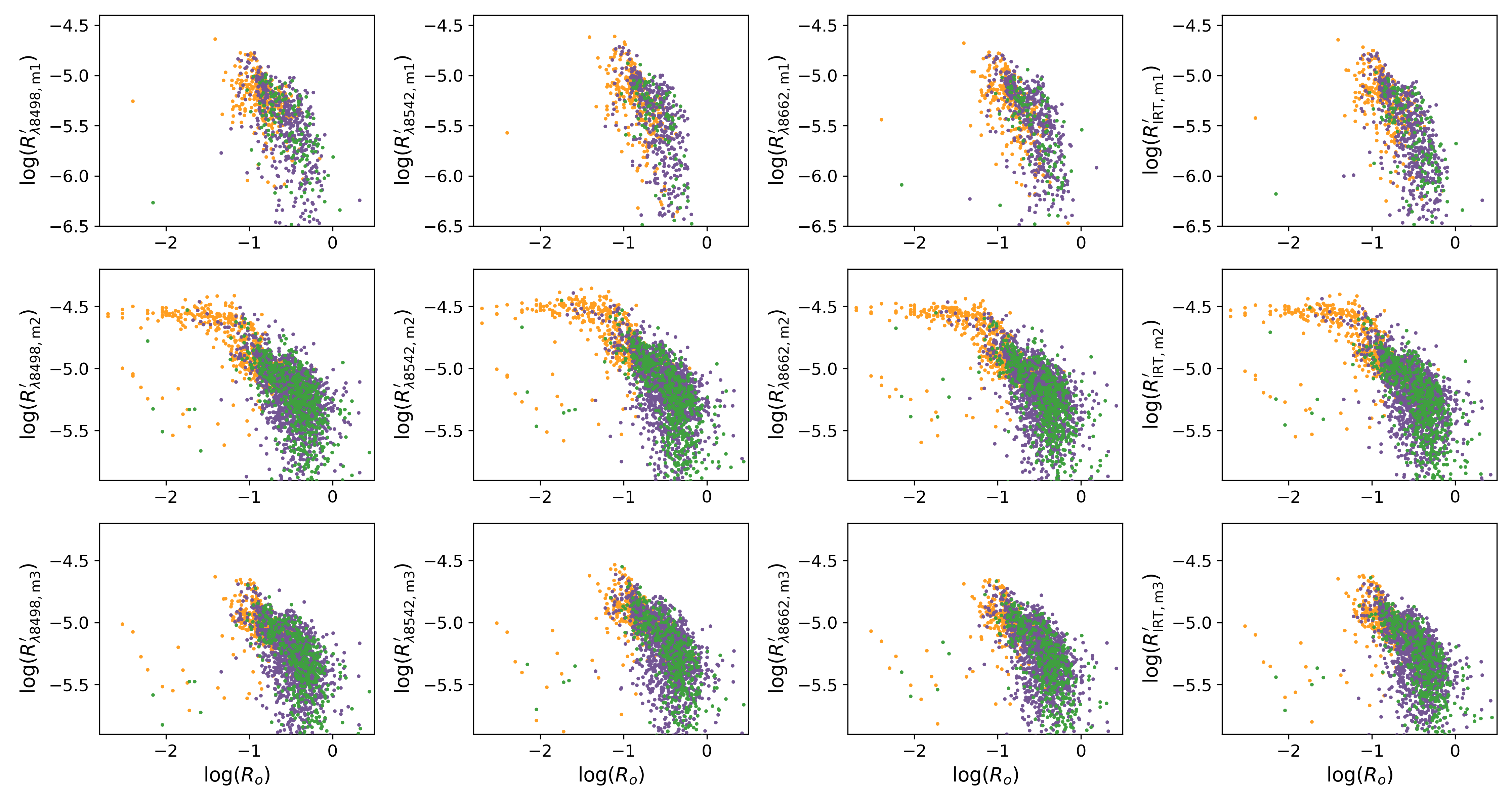}
\caption{Rotation--activity relation constructed with Ca~{\sc ii} IRT activity indices (from Methods 1, 2, and 3). Period measurements were from TESS (orange points), {\it K2} (purple), and {\it Kepler} (green) observations.
\label{rrot.fig}}
\end{center}
\end{figure*}

\begin{figure*}[!htbp]
\begin{center}
\includegraphics[width=0.47\textwidth]{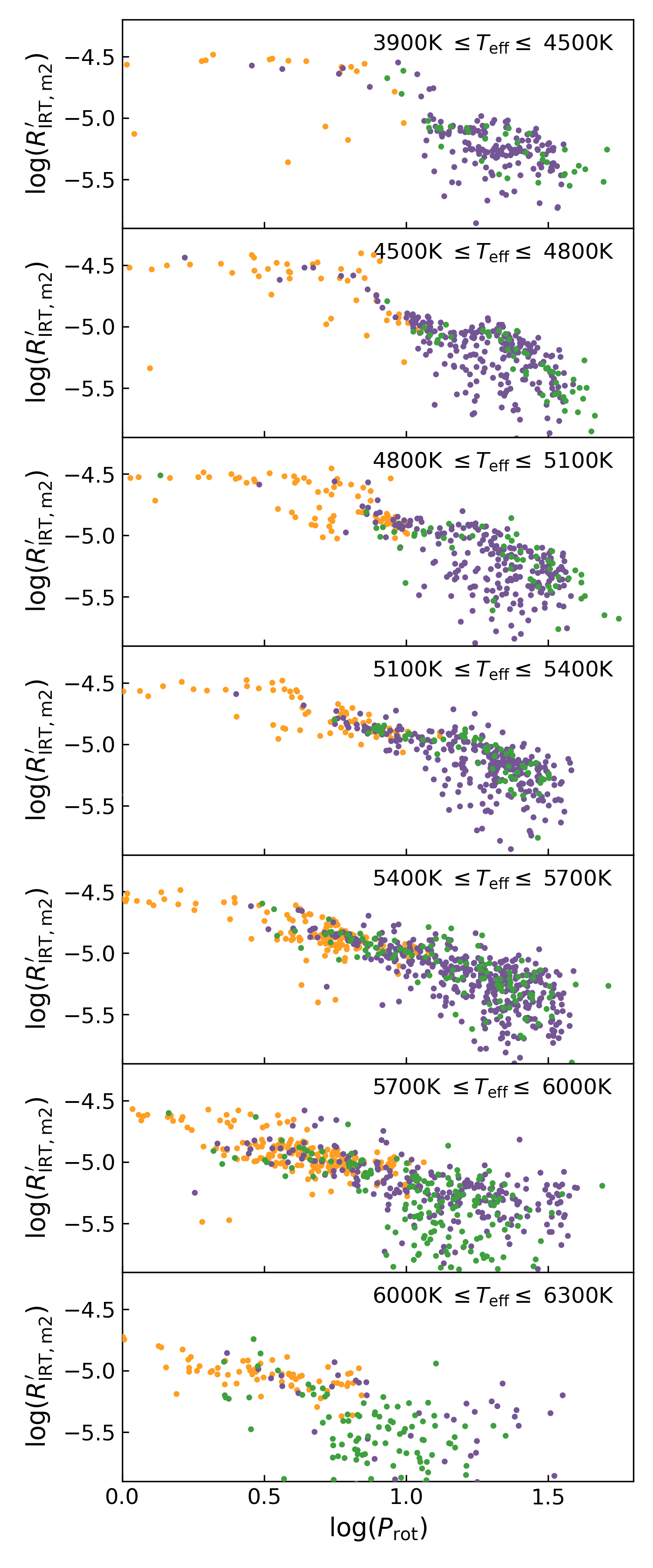}
\includegraphics[width=0.47\textwidth]{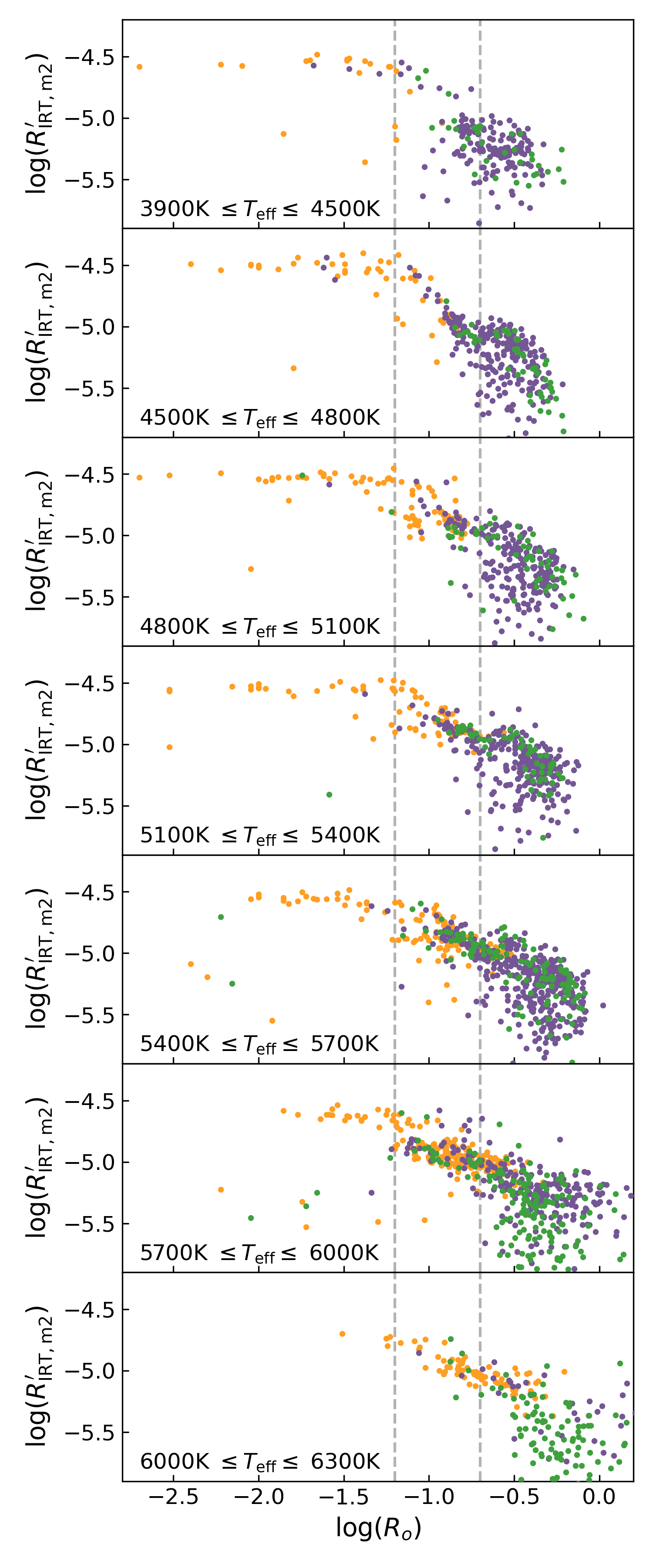}
\caption{
Rotation--activity relation for stars in different $T_{\rm eff}$ bins, with log($P_{\rm rot}$) (left panels) and log($Ro$) (right panels) plotted against log($R'_{\rm IRT, m2}$). Period measurements were from TESS (orange points), {\it K2} (purple), and {\it Kepler} (green) observations.
The two vertical dashed lines mark log($Ro$) $= -$1.2 and $-$0.7.}
\label{rrottype.fig}
\end{center}
\end{figure*}

\subsection{Rotation--activity relation}

Stellar rotation--activity relation is expected to shed light on magnetic dynamos operating in different stars. The canonical rotation--activity relation \citep{2003A&A...397..147P,2011ApJ...743...48W,2016Natur.535..526W} is characterized by two distinct regimes. In the saturated part, rapid rotators maintain a constant activity level, which is thought to be driven by a convective dynamo. In the decay part, slow rotators show activity that declines with decreasing rotation, indicating an interface dynamo.
In this section, we would investigate this relation using the Ca~{\sc ii} IRT activity proxy.

We compiled stellar rotation periods from several catalogs \citep{2014ApJS..211...24M, 2019ApJS..244...21S, 2020A&A...635A..43R, 2025ApJS..276...57G}, spanning observations from the \emph{Kepler}, \emph{K}2, and \emph{TESS}.
When a star was listed in more than one catalog, we gave priority to the period determined from the higher-quality photometry (i.e., \emph{Kepler} $>$ \emph{K}2 $>$ \emph{TESS}). 
To ensure reliable period measurements for the TESS sample  \citep{2025ApJS..276...57G}, we selected only stars with both a high fitting threshold ($R^2 \geq 0.5$) and a high correct classification probability ($CCProb \geq 0.5$).
The Rossby number ($Ro$) is the ratio of the rotation period to the convective turnover time ($\tau_{c}$). Here we used the Yale-Potsdam Stellar Isochrone grids \citep{2017ApJ...838..161S} to calculate global $\tau_{\rm{c}}$, by performing a three-dimensional interpolation over effective temperature, surface gravity, and metallicity for each target.

Figure \ref{rrot.fig} shows the relation between the activity indices ($R'_{\lambda}$, $R'_{\rm IRT}$) and the Rossby number. 
Only Method 2 reveals the saturation regime due to its large sample size. It is worth noting that different methods may yield different decay slopes in the decay regime.
Figure \ref{rrottype.fig} further shows the rotation--activity relation for different spectral types, based on the activity results from Method 2. 
Besides the canonical saturation and decay regimes, fine structures are also visible.
The decay part does not follow a single slope; instead, it exhibits two distinct trends, especially for stars with temperatures between 4500 and 5700\,K. 

In gyrochronology, a main-sequence star evolves through a convective phase (younger, faster rotating, with a convective dynamo) and an interface phase (older, slower rotating, with an interface dynamo), separated by a transitional phase known as the gap \citep{2003ApJ...586..464B,2007ApJ...669.1167B,2010ApJ...721..675B}. Previous studies have linked the saturated and decay regimes in the rotation--activity relation to the convective and interface phases of gyrochronology, respectively \citep{2003ApJ...586L.145B,2011ApJ...743...48W}. 
Recently, \citet{2025A&A...699A.251Y} and \citet{2026ApJ..1000..208H} identified an intermediate regime between the saturated and decay regimes in their rotation--activity relations and associated it with the gap phase, thereby establishing a one-to-one correspondence between the rotation--activity relation and gyrochronology.
Furthermore, \citet{2026ApJ..1000..208H} reported a variable activity decay rate in the intermediate regime, implying a varying core–envelope coupling efficiency during early stellar evolution \citep{2015A&A...584A..30L,2020A&A...636A..76S,2020ApJ...904..140C}.

In Fig. \ref{rrottype.fig}, the decay slope in the range $-1.2 \lesssim {Ro} \lesssim -0.7$ differs from that at $Ro \gtrsim -0.7$.
We suggested the steeper-slope region of $-1.2 \lesssim {Ro} \lesssim -0.7$ corresponds to the gap phase, while the regions before and after this range correspond to the convective and interface sequences, respectively. 
In addition, the shape of the relation closely resembles the findings of \citet{2018A&A...618A..48M}, who found stellar rotation--activity cannot be described by a two-phase model and split their relation into four parts (their Fig. 7). 
Although we did not perform a quantitative fit of our relation, these results suggest the potential of using Ca~{\sc ii} IRT activity, derived with our baseline method, for detailed studies of stellar magnetic dynamos and associated physical processes.

\begin{table*}[ht]
    \centering
    \caption{log$R'_{\lambda}$ and log$R'_{\rm IRT}$ measurements for both dwarfs and giants from Method 1.}
    \setlength{\tabcolsep}{3pt}
    \renewcommand{\arraystretch}{1.2}
    \begin{tabular}{lcccccccc}
    \hline\hline\noalign{\smallskip}
    {\it Gaia} DR3 & $T_{\rm eff}$ & log$g$ & [M/H] & {\tt vbroad} &  log$R^{'}_{\lambda 8498}$ & log$R^{'}_{\lambda 8542}$ & log$R^{'}_{\lambda 8662}$ & log$R'_{\rm IRT}$ \\
    & (K) & & & (km s$^{-1}$) &  & & &  \\
    \hline\noalign{\smallskip}
111600430360448 & 5676$_{-10}^{+14}$ & 3.79$_{-0.01}^{+0.01}$ & $-$0.24$_{-0.08}^{+0.1}$ & 6.9$\pm$8.5 & $-$5.63$\pm$0.09 & ... & $-$5.73$\pm$0.1 & $-$5.68$\pm$0.1\\
133732397478144 & 5784$_{-3}^{+3}$ & 4.29$_{-0.01}^{+0.01}$ & 0.07$_{-0.14}^{+0.15}$ & 12.0$\pm$3.8 & $-$5.7$\pm$0.1 & ... & ... & $-$5.7$\pm$0.1\\
143421843569792 & 5224$_{-11}^{+11}$ & 4.49$_{-0.01}^{+0.01}$ & 0.58$_{-0.08}^{+0.05}$ & 5.2$\pm$6.4 & $-$5.84$\pm$0.23 & $-$6.25$\pm$0.62 & $-$5.7$\pm$0.17 & $-$5.93$\pm$0.4\\
311952064880256 & 5596$_{-8}^{+5}$ & 4.49$_{-0.01}^{+0.01}$ & $-$0.21$_{-0.35}^{+0.44}$ & 10.4$\pm$10.1 & $-$5.44$\pm$0.1 & $-$5.7$\pm$0.19 & $-$5.5$\pm$0.11 & $-$5.55$\pm$0.14\\
397718267299968 & 5054$_{-13}^{+13}$ & 4.53$_{-0.01}^{+0.01}$ & 0.35$_{-0.1}^{+0.15}$ & 15.0$\pm$10.4 & $-$5.72$\pm$0.26 & ... & ... & $-$5.72$\pm$0.26\\
513884247145344 & 5931$_{-30}^{+15}$ & 4.36$_{-0.01}^{+0.01}$ & $-$0.27$_{-0.14}^{+0.11}$ & 8.2$\pm$6.0 & $-$5.36$\pm$0.04 & $-$5.58$\pm$0.08 & $-$5.42$\pm$0.05 & $-$5.45$\pm$0.06\\
1781247492276480 & 5697$_{-22}^{+18}$ & 4.4$_{-0.01}^{+0.01}$ & 0.23$_{-0.07}^{+0.09}$ & 9.1$\pm$9.0 & $-$5.95$\pm$0.21 & ... & ... & $-$5.95$\pm$0.21\\
2662295608063872 & 5804$_{-48}^{+51}$ & 4.48$_{-0.04}^{+0.02}$ & $-$0.11$_{-0.12}^{+0.08}$ & 10.2$\pm$2.8 & $-$5.83$\pm$0.13 & ... & ... & $-$5.83$\pm$0.13\\
2787674293549696 & 5792$_{-10}^{+23}$ & 4.17$_{-0.01}^{+0.01}$ & $-$0.0$_{-0.17}^{+0.22}$ & 12.0$\pm$8.9 & $-$5.63$\pm$0.15 & ... & ... & $-$5.63$\pm$0.15\\
5659667384707456 & 4556$_{-14}^{+10}$ & 4.6$_{-0.01}^{+0.01}$ & 0.02$_{-0.17}^{+0.23}$ & 9.5$\pm$5.6 & $-$5.15$\pm$0.06 & $-$5.08$\pm$0.06 & $-$5.1$\pm$0.05 & $-$5.11$\pm$0.06\\
\noalign{\smallskip}\hline
    \end{tabular}
    \label{m1.tab}\\
\tablefoot{The parameters $T_{\rm eff}$, log$g$, [M/H], and {\tt vbroad} are taken from {\tt teff\_gspphot}, {\tt logg\_gspphot}, {\tt mh\_gspspec}, and {\tt vbroad} in {\it Gaia} DR3 catalog.\\
The entire table is available in machine-readable and Virtual Observatory (VO) forms at CDS. A portion is shown here for guidance regarding its form and content.}
\end{table*}

\begin{table*}[ht]
    \centering
    \caption{log$R'_{\lambda}$ and log$R'_{\rm IRT}$ measurements for dwarfs from Method 2.}
    \setlength{\tabcolsep}{4pt}
    \renewcommand{\arraystretch}{1.2}
    \begin{tabular}{lccccccc}
    \hline\hline\noalign{\smallskip}
    {\it Gaia} DR3 & $T_{\rm eff}$ & log$g$ & [M/H] &  log$R^{'}_{\lambda 8498}$ & log$R^{'}_{\lambda 8542}$ & log$R^{'}_{\lambda 8662}$ & log$R'_{\rm IRT}$ \\
    & (K) & & &  & & &  \\
    \hline\noalign{\smallskip}
9281425163264 & 5909$_{-27}^{+26}$ & 4.03$_{-0.01}^{+0.01}$ & 0.34$_{-0.05}^{+0.06}$ & $-$5.73$\pm$0.08 & $-$5.72$\pm$0.07 & $-$5.76$\pm$0.09 & $-$5.74$\pm$0.08\\
17149805245440 & 5796$_{-65}^{+58}$ & 4.02$_{-0.01}^{+0.01}$ & 0.03$_{-0.27}^{+0.19}$ & $-$5.46$\pm$0.11 & $-$5.43$\pm$0.11 & $-$5.74$\pm$0.22 & $-$5.54$\pm$0.16\\
45664092292480 & 6025$_{-27}^{+27}$ & 4.06$_{-0.02}^{+0.04}$ & $-$0.13$_{-0.14}^{+0.17}$ & $-$5.17$\pm$0.04 & $-$5.36$\pm$0.06 & $-$5.4$\pm$0.06 & $-$5.31$\pm$0.06\\
70476119170048 & 5507$_{-34}^{+37}$ & 4.5$_{-0.01}^{+0.01}$ & $-$0.07$_{-0.16}^{+0.13}$ & $-$5.34$\pm$0.08 & $-$5.4$\pm$0.08 & $-$5.48$\pm$0.08 & $-$5.41$\pm$0.08\\
86105504493056 & 6168$_{-16}^{+20}$ & 4.06$_{-0.01}^{+0.01}$ & $-$0.29$_{-0.08}^{+0.08}$ & $-$5.33$\pm$0.02 & $-$5.42$\pm$0.03 & $-$5.4$\pm$0.03 & $-$5.39$\pm$0.03\\
133732397478144 & 5784$_{-3}^{+3}$ & 4.29$_{-0.01}^{+0.01}$ & 0.07$_{-0.14}^{+0.15}$ & $-$5.23$\pm$0.04 & $-$5.25$\pm$0.04 & $-$5.36$\pm$0.04 & $-$5.28$\pm$0.04\\
142730353190144 & 5718$_{-12}^{+10}$ & 4.42$_{-0.01}^{+0.01}$ & 0.06$_{-0.3}^{+0.07}$ & $-$5.21$\pm$0.03 & $-$5.13$\pm$0.03 & $-$5.17$\pm$0.03 & $-$5.17$\pm$0.03\\
143421843569792 & 5224$_{-11}^{+11}$ & 4.49$_{-0.01}^{+0.01}$ & 0.58$_{-0.08}^{+0.05}$ & $-$5.42$\pm$0.09 & $-$5.26$\pm$0.06 & $-$5.32$\pm$0.07 & $-$5.33$\pm$0.07\\
174959788042496 & 5891$_{-20}^{+21}$ & 4.12$_{-0.01}^{+0.01}$ & 0.15$_{-0.12}^{+0.13}$ & $-$5.26$\pm$0.07 & $-$5.43$\pm$0.09 & $-$5.59$\pm$0.15 & $-$5.42$\pm$0.11\\
180697864082688 & 5812$_{-15}^{+15}$ & 4.25$_{-0.01}^{+0.01}$ & $-$0.24$_{-0.14}^{+0.25}$ & $-$5.24$\pm$0.06 & $-$5.4$\pm$0.07 & $-$5.24$\pm$0.05 & $-$5.29$\pm$0.06\\
\noalign{\smallskip}\hline
    \end{tabular}
    \label{m2.tab}\\
\tablefoot{The parameters $T_{\rm eff}$, log$g$, and [M/H] are taken from {\tt teff\_gspphot}, {\tt logg\_gspphot}, and {\tt mh\_gspspec} in {\it Gaia} DR3 catalog.\\
The entire table is available in machine-readable and Virtual Observatory (VO) forms at CDS. A portion is shown here for guidance regarding its form and content.}
\end{table*}

\begin{table*}[ht]
    \centering
    \caption{log$R'_{\lambda}$ and log$R'_{\rm IRT}$ measurements for dwarfs from Method 3.}
    \setlength{\tabcolsep}{4pt}
    \renewcommand{\arraystretch}{1.2}
    \begin{tabular}{lccccccc}
    \hline\hline\noalign{\smallskip}
{\it Gaia} DR3 & $T_{\rm eff}$ & log$g$ & [M/H] &  log$R^{'}_{\lambda 8498}$ & log$R^{'}_{\lambda 8542}$ & log$R^{'}_{\lambda 8662}$ & log$R'_{\rm IRT}$ \\
    & (K) & & &  & & &  \\
    \hline\noalign{\smallskip}
9281425163264 & 5909$_{-27}^{+26}$ & 4.03$_{-0.01}^{+0.01}$ & 0.34$_{-0.05}^{+0.06}$ & $-$5.67$\pm$0.07 & $-$5.67$\pm$0.06 & $-$5.7$\pm$0.08 & $-$5.68$\pm$0.07\\
12545600306304 & 5250$_{-13}^{+13}$ & 4.46$_{-0.01}^{+0.01}$ & 0.21$_{-0.2}^{+0.13}$ & $-$5.67$\pm$0.24 & ... & $-$6.25$\pm$0.87 & $-$5.96$\pm$0.64\\
17149805245440 & 5796$_{-65}^{+58}$ & 4.02$_{-0.01}^{+0.01}$ & 0.03$_{-0.27}^{+0.19}$ & $-$5.45$\pm$0.11 & $-$5.41$\pm$0.11 & $-$5.72$\pm$0.21 & $-$5.53$\pm$0.15\\
45664092292480 & 6025$_{-27}^{+27}$ & 4.06$_{-0.02}^{+0.04}$ & $-$0.13$_{-0.14}^{+0.17}$ & $-$5.26$\pm$0.05 & $-$5.47$\pm$0.08 & $-$5.55$\pm$0.09 & $-$5.43$\pm$0.07\\
70476119170048 & 5507$_{-34}^{+37}$ & 4.5$_{-0.01}^{+0.01}$ & $-$0.07$_{-0.16}^{+0.13}$ & $-$5.67$\pm$0.18 & $-$5.73$\pm$0.17 & $-$5.96$\pm$0.24 & $-$5.78$\pm$0.2\\
86105504493056 & 6168$_{-16}^{+20}$ & 4.06$_{-0.01}^{+0.01}$ & $-$0.29$_{-0.08}^{+0.08}$ & $-$5.44$\pm$0.03 & $-$5.52$\pm$0.04 & $-$5.51$\pm$0.04 & $-$5.49$\pm$0.04\\
133732397478144 & 5784$_{-3}^{+3}$ & 4.29$_{-0.01}^{+0.01}$ & 0.07$_{-0.14}^{+0.15}$ & $-$5.23$\pm$0.03 & $-$5.24$\pm$0.04 & $-$5.36$\pm$0.04 & $-$5.28$\pm$0.04\\
142730353190144 & 5718$_{-12}^{+10}$ & 4.42$_{-0.01}^{+0.01}$ & 0.06$_{-0.3}^{+0.07}$ & $-$5.21$\pm$0.03 & $-$5.13$\pm$0.03 & $-$5.17$\pm$0.03 & $-$5.17$\pm$0.03\\
143421843569792 & 5224$_{-11}^{+11}$ & 4.49$_{-0.01}^{+0.01}$ & 0.58$_{-0.08}^{+0.05}$ & $-$5.29$\pm$0.06 & $-$5.15$\pm$0.05 & $-$5.2$\pm$0.05 & $-$5.21$\pm$0.06\\
174959788042496 & 5891$_{-20}^{+21}$ & 4.12$_{-0.01}^{+0.01}$ & 0.15$_{-0.12}^{+0.13}$ & $-$5.24$\pm$0.06 & $-$5.4$\pm$0.08 & $-$5.55$\pm$0.14 & $-$5.4$\pm$0.1\\
\noalign{\smallskip}\hline
    \end{tabular}
    \label{m3.tab}\\
\tablefoot{The parameters $T_{\rm eff}$, log$g$, and [M/H] are taken from {\tt teff\_gspphot}, {\tt logg\_gspphot}, and {\tt mh\_gspspec} in {\it Gaia} DR3 catalog.\\
The entire table is available in machine-readable and Virtual Observatory (VO) forms at CDS. A portion is shown here for guidance regarding its form and content.}
\end{table*}

\begin{table*}[ht]
    \centering
    \caption{Linear fit results of log$R'_{\lambda}$ and log$R'_{\rm IRT}$ as a function of log$R_X$: log$R' = a * $log$R_X$ $+$b.}
    \setlength{\tabcolsep}{1pt}
    \begin{tabular}{ccccccccc}
   \hline\hline\noalign{\smallskip}
\multirow{2}{*}{Method} & \multicolumn{2}{c}{$R^{'}_{\lambda 8498}$}  &  \multicolumn{2}{c}{$R^{'}_{\lambda 8542}$} &  \multicolumn{2}{c}{$R^{'}_{\lambda 8662}$}  &  \multicolumn{2}{c}{$R'_{\rm IRT}$} \\
\cmidrule(lr){2-3} \cmidrule(lr){4-5} \cmidrule(lr){6-7} \cmidrule(lr){8-9}
&  $a$  & $b$  &  $a$  & $b$  &  $a$  & $b$  &  $a$  & $b$  \\
\hline\noalign{\smallskip}
m1 &  0.325$\pm$0.013 & $-$3.672$\pm$0.059 & 0.342$\pm$0.015 & $-$3.528$\pm$0.065 & 0.318$\pm$0.014 & $-$3.682$\pm$0.063 & 0.331$\pm$0.014 & $-$3.617$\pm$0.062 \\
m2 & 0.312$\pm$0.003 & $-$3.535$\pm$0.010 & 0.305$\pm$0.003 & $-$3.497$\pm$0.011 & 0.312$\pm$0.002 & $-$3.519$\pm$0.010 & 0.309$\pm$0.002 & $-$3.520$\pm$0.010 \\
m3 & 0.268$\pm$0.008 & $-$3.776$\pm$0.035 & 0.280$\pm$0.008 & $-$3.643$\pm$0.037 & 0.262$\pm$0.008 & $-$3.786$\pm$0.035 & 0.275$\pm$0.008 & $-$3.714$\pm$0.036 \\
\hline\noalign{\smallskip}
    \end{tabular}
    \label{fit.tab}
\end{table*}

\section{Summary}
\label{sum.sec}

In this work, we measured Ca~{\sc ii} IRT activity indices ($R'_{\lambda}$ and $R'_{\rm IRT}$) using published {\it Gaia} RVS spectra with three different methods and compared the results. The procedure involves measuring the $EW$ from the core flux of each Ca~{\sc ii} IRT line, subtracting the photospheric contribution, and converting the excess $EW$ to activity index. The main difference among the three methods lies in how the photospheric contribution is treated. Method 1 follows the standard approach, using PHOENIX templates matched to effective temperature, surface gravity, and metallicity; Method 2 builds a baseline from less active stars as a function of effective temperature alone; Method 3 builds a baseline as a function of both effective temperature and metallicity.

Method 1 has the advantage of providing star-specific photospheric templates based on atmospheric parameters, but it often fails for stars whose observed line cores are deeper than the model predictions, preventing reliable activity estimates.
Methods 2 and 3 are computationally efficient and straightforward to carry out. Method 2 requires only stellar temperature, enabling activity measurements for a large number of stars, although the resulting activity values are subject to metallicity effect.
The limitation of Methods 2 and 3 is that they are less suitable for giant stars, where the large range of log$g$ causes notable variations in line depth and width, making a baseline based solely on temperature and metallicity inadequate for these objects.

Comparing the activity results derived from the three methods, we found that 
Methods 2 and 3 not only return far more objects with valid $R'_{\lambda}$ and $R'_{\rm IRT}$ measurements, but also avoid many very low and unreliable measurements (e.g., log$R'_{\rm IRT}$ $<$ $-$6) seen in Method 1.
In addition, the results from Method 2 are metallicity-dependent: they are lower than those from Method 3 for metal-rich stars, but higher for metal-poor ones.

We examined the relationship between the Ca~{\sc ii} IRT activity indices and other activity indicators, including the Ca H\&K emission ($R'_{\rm HK}$) and X-ray emission ($R_X$). In general, the Ca~{\sc ii} IRT indices correlate positively with these proxies. Due to the larger sample size, Method 2 yields the widest range of activity levels and the tightest correlations.

The correlation between Ca~{\sc ii} IRT activity and stellar rotation further confirms that $R'_{\lambda}$ and $R'_{\rm IRT}$ are reliable activity indicators. Clear saturation and decay regimes can be seen in the rotation--activity relation, particularly for the results from Method 2.
Of particular interest is the presence of fine structures within the relation. The decay part appears to split into two regions: one in the range $-1.2 \lesssim {\rm Ro} \lesssim -0.7$ and the other at $\rm Ro \gtrsim -0.7$.
The former interval may correspond to the gap regime defined in gyrochronology. 

A difference between our method and the canonical method is that our method not only estimates the photospheric contribution but also accounts for weak chromospheric emission unrelated to magnetic activity, such as acoustic-wave heating \citep{2025A&A...699A..25U}. However, one caveat is that for stars at low activity levels, the photospheric and non-magnetic chromospheric contributions complicate the physical interpretation and can lead to large uncertainty in activity calculation. Therefore, one should be cautious when applying any method to low-activity stars.

To date, fewer than 10,000 stars have high-resolution spectroscopic observations of Ca~{\sc ii} H\&K lines \citep[e.g.,][]{2010ApJ...725..875I,2018A&A...616A.108B,2024ApJS..274...35I}. 
Considering that the activity indices calculated from spectra are strongly affected by spectral resolution, the vast sample of high-resolution Ca~{\sc ii} IRT spectra from {\it Gaia} offers an invaluable database for studying chromospheric activity. 
Our results suggest that until the template-based methods can provide a reliable estimate of the photospheric contribution to the Ca~{\sc ii} IRT lines, the baseline method can serve as a robust alternative for deriving accurate activity estimates.
This approach will be especially useful for deriving Ca~{\sc ii} IRT activity indices from the upcoming {\it Gaia} DR4 data.

\section*{Data availability}
Tables 1, 2, and 3 are only available in electronic form at the CDS via anonymous ftp to \url{cdsarc.u-strasbg.fr} (130.79.128.5) or via \url{http://cdsweb.u-strasbg.fr/cgi-bin/qcat?J/A+A/}.

\begin{acknowledgements}

This work was supported by the National Natural Science Foundation of China (NSFC; grant Nos. 12588202/12673041/12273057), the National Key R\&D Program of China (grant Nos. 2025YFF0511203/2025YFF0511200), the Strategic Priority Research Program of the Chinese Academy of Sciences (grant No. XDB1160302), the NAOC Project with No. E6517401, Science Research Grants from the China Manned Space Project with No. CMS-CSST-2021-A08, and the New Cornerstone Science Foundation through the New Cornerstone Investigator Program and the XPLORER PRIZE.
This work presents results from the European Space Agency (ESA) space mission {\it Gaia}. {\it Gaia} data are being processed by the {\it Gaia} Data Processing and Analysis Consortium (DPAC). This research has made use of the SIMBAD database, operated at CDS, Strasbourg, France.

\end{acknowledgements}

\bibliography{paper.bib}{}

@ARTICLE{2026ApJS..287...38Y,
       author = {{Yang}, Xiaozhen and {Fu}, Xiaoting and {Jian}, Mingjie and {Zhao}, Jingkun and {Yuan}, Hailong and {Bai}, Zhongrui and {Wang}, Mengxin and {Dong}, Yiqiao and {Yang}, Mingkuan and {Jiang}, Ziyue and et al.},
        title = "{Revisiting Ca II Activity Indices in FGK Stars: Systematic Biases in Infrared Triplet Measurements}",
      journal = {\apjs},
         year = 2026,
        month = nov,
       volume = {287},
       number = {1},
          eid = {38},
        pages = {38},
          doi = {10.3847/1538-4365/aeb011},
       adsurl = {https://ui.adsabs.harvard.edu/abs/2026ApJS..287...38Y}
}

@ARTICLE{2025A&A...699A..25U,
       author = {{Udn{\ae}s}, Elias R. and {Pereira}, Tiago M.~D.},
        title = "{Characteristics of acoustic-wave heating in simulations of the quiet Sun chromosphere}",
      journal = {\aap},
         year = 2025,
        month = jul,
       volume = {699},
          eid = {A25},
        pages = {A25},
          doi = {10.1051/0004-6361/202453028},
archivePrefix = {arXiv},
       eprint = {2505.21047},
 primaryClass = {astro-ph.SR},
       adsurl = {https://ui.adsabs.harvard.edu/abs/2025A&A...699A..25U}
}

@ARTICLE{2009A&ARv..17..251S,
       author = {{Strassmeier}, Klaus G.},
        title = "{Starspots}",
      journal = {\aapr},
         year = 2009,
        month = sep,
       volume = {17},
       number = {3},
        pages = {251-308},
          doi = {10.1007/s00159-009-0020-6},
       adsurl = {https://ui.adsabs.harvard.edu/abs/2009A&ARv..17..251S}
}

@ARTICLE{2013MNRAS.431.2063S,
       author = {{Stelzer}, B. and {Marino}, A. and {Micela}, G. and {L{\'o}pez-Santiago}, J. and {Liefke}, C.},
        title = "{The UV and X-ray activity of the M dwarfs within 10 pc of the Sun}",
      journal = {\mnras},
         year = 2013,
        month = may,
       volume = {431},
       number = {3},
        pages = {2063-2079},
          doi = {10.1093/mnras/stt225},
archivePrefix = {arXiv},
       eprint = {1302.1061},
 primaryClass = {astro-ph.SR},
       adsurl = {https://ui.adsabs.harvard.edu/abs/2013MNRAS.431.2063S}
}

@ARTICLE{2001AJ....122.3466M,
       author = {{Mason}, Brian D. and {Wycoff}, Gary L. and {Hartkopf}, William I. and {Douglass}, Geoffrey G. and {Worley}, Charles E.},
        title = "{The 2001 US Naval Observatory Double Star CD-ROM. I. The Washington Double Star Catalog}",
      journal = {\aj},
         year = 2001,
        month = dec,
       volume = {122},
       number = {6},
        pages = {3466-3471},
          doi = {10.1086/323920},
       adsurl = {https://ui.adsabs.harvard.edu/abs/2001AJ....122.3466M}
}

@ARTICLE{2026ApJS..284...75B,
       author = {{Boyle}, Andrew W. and {Bouma}, Luke G. and {Mann}, Andrew W.},
        title = "{The TESS All-Sky Rotation Survey: Periods for 1,046,317 Stars within 500 pc}",
      journal = {\apjs},
         year = 2026,
        month = jun,
       volume = {284},
       number = {2},
          eid = {75},
        pages = {75},
          doi = {10.3847/1538-4365/ae6657},
archivePrefix = {arXiv},
       eprint = {2603.05586},
 primaryClass = {astro-ph.SR},
       adsurl = {https://ui.adsabs.harvard.edu/abs/2026ApJS..284...75B}
}

@ARTICLE{2003ApJ...586..464B,
       author = {{Barnes}, Sydney A.},
        title = "{On the Rotational Evolution of Solar- and Late-Type Stars, Its Magnetic Origins, and the Possibility of Stellar Gyrochronology}",
      journal = {\apj},
         year = 2003,
        month = mar,
       volume = {586},
       number = {1},
        pages = {464-479},
          doi = {10.1086/367639},
archivePrefix = {arXiv},
       eprint = {astro-ph/0303631},
 primaryClass = {astro-ph},
       adsurl = {https://ui.adsabs.harvard.edu/abs/2003ApJ...586..464B}
}

@ARTICLE{2007ApJ...669.1167B,
       author = {{Barnes}, Sydney A.},
        title = "{Ages for Illustrative Field Stars Using Gyrochronology: Viability, Limitations, and Errors}",
      journal = {\apj},
         year = 2007,
        month = nov,
       volume = {669},
       number = {2},
        pages = {1167-1189},
          doi = {10.1086/519295},
archivePrefix = {arXiv},
       eprint = {0704.3068},
 primaryClass = {astro-ph},
       adsurl = {https://ui.adsabs.harvard.edu/abs/2007ApJ...669.1167B}
}

@ARTICLE{2010ApJ...721..675B,
       author = {{Barnes}, Sydney A. and {Kim}, Yong-Cheol},
        title = "{Angular Momentum Loss from Cool Stars: An Empirical Expression and Connection to Stellar Activity}",
      journal = {\apj},
         year = 2010,
        month = sep,
       volume = {721},
       number = {1},
        pages = {675-685},
          doi = {10.1088/0004-637X/721/1/675},
archivePrefix = {arXiv},
       eprint = {1104.2350},
 primaryClass = {astro-ph.SR},
       adsurl = {https://ui.adsabs.harvard.edu/abs/2010ApJ...721..675B}
}

@ARTICLE{2003ApJ...586L.145B,
       author = {{Barnes}, Sydney A.},
        title = "{A Connection between the Morphology of the X-Ray Emission and Rotation for Solar-Type Stars in Open Clusters}",
      journal = {\apjl},
         year = 2003,
        month = apr,
       volume = {586},
       number = {2},
        pages = {L145-L147},
          doi = {10.1086/374681},
archivePrefix = {arXiv},
       eprint = {astro-ph/0303632},
 primaryClass = {astro-ph},
       adsurl = {https://ui.adsabs.harvard.edu/abs/2003ApJ...586L.145B}
}

@ARTICLE{2020A&A...636A..76S,
       author = {{Spada}, F. and {Lanzafame}, A.~C.},
        title = "{Competing effect of wind braking and interior coupling in the rotational evolution of solar-like stars}",
      journal = {\aap},
         year = 2020,
        month = apr,
       volume = {636},
          eid = {A76},
        pages = {A76},
          doi = {10.1051/0004-6361/201936384},
archivePrefix = {arXiv},
       eprint = {1908.00345},
 primaryClass = {astro-ph.SR},
       adsurl = {https://ui.adsabs.harvard.edu/abs/2020A&A...636A..76S}
}

@ARTICLE{2017ApJ...838..161S,
       author = {{Spada}, F. and {Demarque}, P. and {Kim}, Y. -C. and {Boyajian}, T.~S. and {Brewer}, J.~M.},
        title = "{The Yale-Potsdam Stellar Isochrones}",
      journal = {\apj},
         year = 2017,
        month = apr,
       volume = {838},
       number = {2},
          eid = {161},
        pages = {161},
          doi = {10.3847/1538-4357/aa661d},
archivePrefix = {arXiv},
       eprint = {1703.03975},
 primaryClass = {astro-ph.SR},
       adsurl = {https://ui.adsabs.harvard.edu/abs/2017ApJ...838..161S}
}

@ARTICLE{2020ApJ...904..140C,
       author = {{Curtis}, Jason Lee and {Ag{\"u}eros}, Marcel A. and {Matt}, Sean P. and {Covey}, Kevin R. and {Douglas}, Stephanie T. and {Angus}, Ruth and {Saar}, Steven H. and {Cody}, Ann Marie and {Vanderburg}, Andrew and {Law}, Nicholas M. and {Kraus}, Adam L. and {Latham}, David W. and {Baranec}, Christoph and {Riddle}, Reed and {Ziegler}, Carl and {Lund}, Mikkel N. and {Torres}, Guillermo and {Meibom}, S{\o}ren and {Aguirre}, Victor Silva and {Wright}, Jason T.},
        title = "{When Do Stalled Stars Resume Spinning Down? Advancing Gyrochronology with Ruprecht 147}",
      journal = {\apj},
         year = 2020,
        month = dec,
       volume = {904},
       number = {2},
          eid = {140},
        pages = {140},
          doi = {10.3847/1538-4357/abbf58},
archivePrefix = {arXiv},
       eprint = {2010.02272},
 primaryClass = {astro-ph.SR},
       adsurl = {https://ui.adsabs.harvard.edu/abs/2020ApJ...904..140C}
}

@ARTICLE{2015A&A...584A..30L,
       author = {{Lanzafame}, A.~C. and {Spada}, F.},
        title = "{Rotational evolution of slow-rotator sequence stars}",
      journal = {\aap},
         year = 2015,
        month = dec,
       volume = {584},
          eid = {A30},
        pages = {A30},
          doi = {10.1051/0004-6361/201526770},
archivePrefix = {arXiv},
       eprint = {1506.05298},
 primaryClass = {astro-ph.SR},
       adsurl = {https://ui.adsabs.harvard.edu/abs/2015A&A...584A..30L}
}

@ARTICLE{2026ApJS..286...71L,
       author = {{Li}, Xue and {Wang}, Song and {Han}, Henggeng and {Ma}, Jun and {Huang}, Yang and {Liu}, Jifeng},
        title = "{Evolution of Stellar Activity and Habitable Zone. III. X-Ray Activity of Dwarfs in Open Clusters and Field Stars}",
      journal = {\apjs},
         year = 2026,
        month = oct,
       volume = {286},
       number = {2},
          eid = {71},
        pages = {71},
          doi = {10.3847/1538-4365/aea077},
archivePrefix = {arXiv},
       eprint = {2608.29248},
 primaryClass = {astro-ph.SR},
       adsurl = {https://ui.adsabs.harvard.edu/abs/2026ApJS..286...71L}
}

@ARTICLE{2004A&ARv..12...71G,
       author = {{G{\"u}del}, Manuel},
        title = "{X-ray astronomy of stellar coronae}",
      journal = {\aapr},
         year = 2004,
        month = sep,
       volume = {12},
       number = {2-3},
        pages = {71-237},
          doi = {10.1007/s00159-004-0023-2},
archivePrefix = {arXiv},
       eprint = {astro-ph/0406661},
 primaryClass = {astro-ph},
       adsurl = {https://ui.adsabs.harvard.edu/abs/2004A&ARv..12...71G}
}

@ARTICLE{2018MNRAS.477.3145J,
       author = {{Jayasinghe}, T. and {Kochanek}, C.~S. and {Stanek}, K.~Z. and {Shappee}, B.~J. and {Holoien}, T.~W.-S. and {Thompson}, Todd A. and {Prieto}, J.~L. and {Dong}, Subo and {Pawlak}, M. and {Shields}, J.~V. and et al.},
        title = "{The ASAS-SN catalogue of variable stars I: The Serendipitous Survey}",
      journal = {\mnras},
         year = 2018,
        month = jul,
       volume = {477},
       number = {3},
        pages = {3145-3163},
          doi = {10.1093/mnras/sty838},
archivePrefix = {arXiv},
       eprint = {1803.01001},
 primaryClass = {astro-ph.SR},
       adsurl = {https://ui.adsabs.harvard.edu/abs/2018MNRAS.477.3145J}
}

@ARTICLE{2000A&AS..143....9W,
       author = {{Wenger}, M. and {Ochsenbein}, F. and {Egret}, D. and {Dubois}, P. and {Bonnarel}, F. and {Borde}, S. and {Genova}, F. and {Jasniewicz}, G. and {Lalo{\"e}}, S. and {Lesteven}, S. and et al.},
        title = "{The SIMBAD astronomical database. The CDS reference database for astronomical objects}",
      journal = {\aaps},
         year = 2000,
        month = apr,
       volume = {143},
        pages = {9-22},
          doi = {10.1051/aas:2000332},
archivePrefix = {arXiv},
       eprint = {astro-ph/0002110},
 primaryClass = {astro-ph},
       adsurl = {https://ui.adsabs.harvard.edu/abs/2000A&AS..143....9W}
}

@ARTICLE{2004A&A...424..727P,
       author = {{Pourbaix}, D. and {Tokovinin}, A.~A. and {Batten}, A.~H. and {Fekel}, F.~C. and {Hartkopf}, W.~I. and {Levato}, H. and {Morrell}, N.~I. and {Torres}, G. and {Udry}, S.},
        title = "{S$_{B$^{9}$}$: The ninth catalogue of spectroscopic binary orbits}",
      journal = {\aap},
         year = 2004,
        month = sep,
       volume = {424},
        pages = {727-732},
          doi = {10.1051/0004-6361:20041213},
archivePrefix = {arXiv},
       eprint = {astro-ph/0406573},
 primaryClass = {astro-ph},
       adsurl = {https://ui.adsabs.harvard.edu/abs/2004A&A...424..727P}
}

@ARTICLE{2019A&A...623A..24F,
       author = {{Fuhrmeister}, B. and {Czesla}, S. and {Schmitt}, J.~H.~M.~M. and {Johnson}, E.~N. and {Sch{\"o}fer}, P. and {Jeffers}, S.~V. and {Caballero}, J.~A. and {Zechmeister}, M. and {Reiners}, A. and {Ribas}, I. and et al.},
        title = "{The CARMENES search for exoplanets around M dwarfs. Period search in H{\ensuremath{\alpha}}, Na I D, and Ca II IRT lines}",
      journal = {\aap},
         year = 2019,
        month = mar,
       volume = {623},
          eid = {A24},
        pages = {A24},
          doi = {10.1051/0004-6361/201834483},
archivePrefix = {arXiv},
       eprint = {1901.05173},
 primaryClass = {astro-ph.SR},
       adsurl = {https://ui.adsabs.harvard.edu/abs/2019A&A...623A..24F}
}

@ARTICLE{2008A&A...479..557F,
       author = {{Frasca}, A. and {Biazzo}, K. and {Ta{\textcommabelow s}}, G. and {Evren}, S. and {Lanzafame}, A.~C.},
        title = "{Spots, plages, and flares on {\ensuremath{\lambda}} Andromedae and II Pegasi}",
      journal = {\aap},
         year = 2008,
        month = feb,
       volume = {479},
       number = {2},
        pages = {557-565},
          doi = {10.1051/0004-6361:20077915},
archivePrefix = {arXiv},
       eprint = {0711.3322},
 primaryClass = {astro-ph},
       adsurl = {https://ui.adsabs.harvard.edu/abs/2008A&A...479..557F}
}

@ARTICLE{2024ApJS..272....6H,
       author = {{Huang}, Xin and {He}, Yuji and {Bai}, Zhongrui and {Yuan}, Hailong and {Yang}, Mingkuan and {Zhou}, Ming and {Dong}, Yiqiao and {Wang}, Mengxin and {He}, Han and {Zhang}, Jinghua and et al.},
        title = "{The Near-infrared Ca II Triplet as a Stellar Activity Indicator: A Library and Comparative Study}",
      journal = {\apjs},
         year = 2024,
        month = may,
       volume = {272},
       number = {1},
          eid = {6},
        pages = {6},
          doi = {10.3847/1538-4365/ad33bc},
archivePrefix = {arXiv},
       eprint = {2311.03120},
 primaryClass = {astro-ph.SR},
       adsurl = {https://ui.adsabs.harvard.edu/abs/2024ApJS..272....6H}
}

@ARTICLE{1993ApJS...86..293D,
       author = {{Dempsey}, Robert C. and {Bopp}, Bernard W. and {Henry}, Gregory W. and {Hall}, Douglas S.},
        title = "{Observations of the CA II Infrared Triplet in Chromospherically Active Single and Binary Stars}",
      journal = {\apjs},
         year = 1993,
        month = may,
       volume = {86},
        pages = {293},
          doi = {10.1086/191779},
       adsurl = {https://ui.adsabs.harvard.edu/abs/1993ApJS...86..293D}
}

@ARTICLE{2018A&A...616A.108B,
       author = {{Boro Saikia}, S. and {Marvin}, C.~J. and {Jeffers}, S.~V. and {Reiners}, A. and {Cameron}, R. and {Marsden}, S.~C. and {Petit}, P. and {Warnecke}, J. and {Yadav}, A.~P.},
        title = "{Chromospheric activity catalogue of 4454 cool stars. Questioning the active branch of stellar activity cycles}",
      journal = {\aap},
         year = 2018,
        month = aug,
       volume = {616},
          eid = {A108},
        pages = {A108},
          doi = {10.1051/0004-6361/201629518},
archivePrefix = {arXiv},
       eprint = {1803.11123},
 primaryClass = {astro-ph.SR},
       adsurl = {https://ui.adsabs.harvard.edu/abs/2018A&A...616A.108B}
}

@ARTICLE{2007A&A...466.1089B,
       author = {{Bus{\`a}}, I. and {Aznar Cuadrado}, R. and {Terranegra}, L. and {Andretta}, V. and {Gomez}, M.~T.},
        title = "{The Ca II infrared triplet as a stellar activity diagnostic. II. Test and calibration with high resolution observations}",
      journal = {\aap},
         year = 2007,
        month = may,
       volume = {466},
       number = {3},
        pages = {1089-1098},
          doi = {10.1051/0004-6361:20065588},
       adsurl = {https://ui.adsabs.harvard.edu/abs/2007A&A...466.1089B}
}

@ARTICLE{2017A&A...605A.113M,
       author = {{Martin}, J. and {Fuhrmeister}, B. and {Mittag}, M. and {Schmidt}, T.~O.~B. and {Hempelmann}, A. and {Gonz{\'a}lez-P{\'e}rez}, J.~N. and {Schmitt}, J.~H.~M.~M.},
        title = "{The Ca II infrared triplet's performance as an activity indicator compared to Ca II H and K. Empirical relations to convert Ca II infrared triplet measurements to common activity indices}",
      journal = {\aap},
         year = 2017,
        month = sep,
       volume = {605},
          eid = {A113},
        pages = {A113},
          doi = {10.1051/0004-6361/201630298},
       adsurl = {https://ui.adsabs.harvard.edu/abs/2017A&A...605A.113M}
}

@ARTICLE{2005A&A...430..669A,
       author = {{Andretta}, V. and {Bus{\`a}}, I. and {Gomez}, M.~T. and {Terranegra}, L.},
        title = "{The Ca II Infrared Triplet as a stellar activity diagnostic . I. Non-LTE photospheric profiles and definition of the R$_{IRT}$ indicator}",
      journal = {\aap},
         year = 2005,
        month = feb,
       volume = {430},
        pages = {669-677},
          doi = {10.1051/0004-6361:20041745},
       adsurl = {https://ui.adsabs.harvard.edu/abs/2005A&A...430..669A}
}

@ARTICLE{2025A&A...697A.230F,
       author = {{Freund}, S. and {Czesla}, S. and {Fuhrmeister}, B. and {Predehl}, P. and {Robrade}, J. and {Schneider}, P.~C. and {Schmitt}, J.~H.~M.~M.},
        title = "{The stellar corona-chromosphere connection: A comprehensive study of X-ray and Ca II IRT fluxes from eROSITA and Gaia}",
      journal = {\aap},
         year = 2025,
        month = may,
       volume = {697},
          eid = {A230},
        pages = {A230},
          doi = {10.1051/0004-6361/202451421},
archivePrefix = {arXiv},
       eprint = {2504.17593},
 primaryClass = {astro-ph.SR},
       adsurl = {https://ui.adsabs.harvard.edu/abs/2025A&A...697A.230F}
}

@ARTICLE{2023A&A...674A..30L,
       author = {{Lanzafame}, A.~C. and {Brugaletta}, E. and {Fr{\'e}mat}, Y. and {Sordo}, R. and {Creevey}, O.~L. and {Andretta}, V. and {Scandariato}, G. and {Bus{\`a}}, I. and {Distefano}, E. and {Korn}, A.~J. and et al.},
        title = "{Gaia Data Release 3. Stellar chromospheric activity and mass accretion from Ca II IRT observed by the Radial Velocity Spectrometer}",
      journal = {\aap},
         year = 2023,
        month = jun,
       volume = {674},
          eid = {A30},
        pages = {A30},
          doi = {10.1051/0004-6361/202244156},
archivePrefix = {arXiv},
       eprint = {2206.05766},
 primaryClass = {astro-ph.SR},
       adsurl = {https://ui.adsabs.harvard.edu/abs/2023A&A...674A..30L}
}

@ARTICLE{2023A&A...674A...6S,
       author = {{Sartoretti}, P. and {Marchal}, O. and {Babusiaux}, C. and {Jordi}, C. and {Guerrier}, A. and {Panuzzo}, P. and {Katz}, D. and {Seabroke}, G.~M. and {Th{\'e}venin}, F. and {Cropper}, M. and et al.},
        title = "{Gaia Data Release 3. G$_{RVS}$ photometry from the RVS spectra}",
      journal = {\aap},
         year = 2023,
        month = jun,
       volume = {674},
          eid = {A6},
        pages = {A6},
          doi = {10.1051/0004-6361/202243615},
archivePrefix = {arXiv},
       eprint = {2206.05725},
 primaryClass = {astro-ph.GA},
       adsurl = {https://ui.adsabs.harvard.edu/abs/2023A&A...674A...6S}
}

@ARTICLE{1991ApJ...375..722S,
       author = {{Soderblom}, David R. and {Duncan}, Douglas K. and {Johnson}, Dean R.~H.},
        title = "{The Chromospheric Emission--Age Relation for Stars of the Lower Main Sequence and Its Implications for the Star Formation Rate}",
      journal = {\apj},
         year = 1991,
        month = jul,
       volume = {375},
        pages = {722},
          doi = {10.1086/170238},
       adsurl = {https://ui.adsabs.harvard.edu/abs/1991ApJ...375..722S}
}

@ARTICLE{2025ApJ...983L..31C,
       author = {{Carvalho-Silva}, Gabriela and {Mel{\'e}ndez}, Jorge and {Rathsam}, Anne and {Shejeelammal}, J. and {Martos}, Giulia and {Lorenzo-Oliveira}, Diego and {Spina}, Lorenzo and {Ribeiro Alves}, D{\'e}bora},
        title = "{A New Age─Activity Relation For Solar Analogs that Accounts for Metallicity}",
      journal = {\apjl},
         year = 2025,
        month = apr,
       volume = {983},
       number = {2},
          eid = {L31},
        pages = {L31},
          doi = {10.3847/2041-8213/adc382},
archivePrefix = {arXiv},
       eprint = {2504.17482},
 primaryClass = {astro-ph.SR},
       adsurl = {https://ui.adsabs.harvard.edu/abs/2025ApJ...983L..31C}
}

@ARTICLE{2018A&A...619A..73L,
       author = {{Lorenzo-Oliveira}, Diego and {Freitas}, Fabr{\'\i}cio C. and {Mel{\'e}ndez}, Jorge and {Bedell}, Megan and {Ram{\'\i}rez}, Iv{\'a}n and {Bean}, Jacob L. and {Asplund}, Martin and {Spina}, Lorenzo and {Dreizler}, Stefan and {Alves-Brito}, Alan and {Casagrande}, Luca},
        title = "{The Solar Twin Planet Search. The age-chromospheric activity relation}",
      journal = {\aap},
         year = 2018,
        month = nov,
       volume = {619},
          eid = {A73},
        pages = {A73},
          doi = {10.1051/0004-6361/201629294},
archivePrefix = {arXiv},
       eprint = {1806.08014},
 primaryClass = {astro-ph.SR},
       adsurl = {https://ui.adsabs.harvard.edu/abs/2018A&A...619A..73L}
}

@ARTICLE{2025ApJ...984....2H,
       author = {{Han}, Henggeng and {Wang}, Song and {Li}, Xue and {Zheng}, Chuanjie and {Liu}, Jifeng},
        title = "{Impact of Spectral Resolution on S-index and Its Application to Spectroscopic Surveys}",
      journal = {\apj},
         year = 2025,
        month = may,
       volume = {984},
       number = {1},
          eid = {2},
        pages = {2},
          doi = {10.3847/1538-4357/adc600},
archivePrefix = {arXiv},
       eprint = {2503.20165},
 primaryClass = {astro-ph.SR},
       adsurl = {https://ui.adsabs.harvard.edu/abs/2025ApJ...984....2H}
}

@ARTICLE{2024ApJS..273....8H,
       author = {{Han}, Henggeng and {Wang}, Song and {Zheng}, Chuanjie and {Li}, Xue and {Xiao}, Kai and {Liu}, Jifeng},
        title = "{Stellar X-Ray Activity and Habitability Revealed by the ROSAT Sky Survey}",
      journal = {\apjs},
         year = 2024,
        month = jul,
       volume = {273},
       number = {1},
          eid = {8},
        pages = {8},
          doi = {10.3847/1538-4365/ad4b17},
archivePrefix = {arXiv},
       eprint = {2405.02863},
 primaryClass = {astro-ph.SR},
       adsurl = {https://ui.adsabs.harvard.edu/abs/2024ApJS..273....8H}
}

@ARTICLE{2010ApJ...725..875I,
       author = {{Isaacson}, Howard and {Fischer}, Debra},
        title = "{Chromospheric Activity and Jitter Measurements for 2630 Stars on the California Planet Search}",
      journal = {\apj},
         year = 2010,
        month = dec,
       volume = {725},
       number = {1},
        pages = {875-885},
          doi = {10.1088/0004-637X/725/1/875},
archivePrefix = {arXiv},
       eprint = {1009.2301},
 primaryClass = {astro-ph.EP},
       adsurl = {https://ui.adsabs.harvard.edu/abs/2010ApJ...725..875I}
}

@ARTICLE{2024ApJS..274...35I,
       author = {{Isaacson}, Howard and {Howard}, Andrew W. and {Fulton}, Benjamin and {Petigura}, Erik A. and {Weiss}, Lauren M. and {Kane}, Stephen R. and {Carter}, Brad and {Beard}, Corey and {Giacalone}, Steven and {Van Zandt}, Judah and et al.},
        title = "{The California Legacy Survey. V. Chromospheric Activity Cycles in Main-sequence Stars}",
      journal = {\apjs},
         year = 2024,
        month = oct,
       volume = {274},
       number = {2},
          eid = {35},
        pages = {35},
          doi = {10.3847/1538-4365/ad676c},
archivePrefix = {arXiv},
       eprint = {2406.17332},
 primaryClass = {astro-ph.SR},
       adsurl = {https://ui.adsabs.harvard.edu/abs/2024ApJS..274...35I}
}

@ARTICLE{2025AJ....170..212S,
       author = {{Sayeed}, Maryum and {Huber}, Daniel and {Chontos}, Ashley and {Li}, Yaguang},
        title = "{A Homogeneous Catalog of Oscillating Solar-type Stars Observed by the Kepler Mission and a New Amplitude Scaling Relation Including Chromospheric Activity}",
      journal = {\aj},
         year = 2025,
        month = oct,
       volume = {170},
       number = {4},
          eid = {212},
        pages = {212},
          doi = {10.3847/1538-3881/adf648},
archivePrefix = {arXiv},
       eprint = {2503.15599},
 primaryClass = {astro-ph.SR},
       adsurl = {https://ui.adsabs.harvard.edu/abs/2025AJ....170..212S}
}

@ARTICLE{2000A&AS..146..103M,
       author = {{Montes}, D. and {Fern{\'a}ndez-Figueroa}, M.~J. and {De Castro}, E. and {Cornide}, M. and {Latorre}, A. and {Sanz-Forcada}, J.},
        title = "{Multiwavelength optical observations of chromospherically active binary systems. III. High resolution echelle spectra from Ca II H \& K to Ca II IRT}",
      journal = {\aaps},
         year = 2000,
        month = oct,
       volume = {146},
        pages = {103-140},
          doi = {10.1051/aas:2000359},
archivePrefix = {arXiv},
       eprint = {astro-ph/0007041},
 primaryClass = {astro-ph},
       adsurl = {https://ui.adsabs.harvard.edu/abs/2000A&AS..146..103M}
}

@ARTICLE{2000A&A...353..666C,
       author = {{Chmielewski}, Yves},
        title = "{The infrared triplet lines of ionized calcium as a diagnostic tool for F, G, K-type stellar atmospheres}",
      journal = {\aap},
         year = 2000,
        month = jan,
       volume = {353},
        pages = {666-690},
       adsurl = {https://ui.adsabs.harvard.edu/abs/2000A&A...353..666C}
}

@ARTICLE{2011ApJ...743...48W,
       author = {{Wright}, Nicholas J. and {Drake}, Jeremy J. and {Mamajek}, Eric E. and {Henry}, Gregory W.},
        title = "{The Stellar-activity-Rotation Relationship and the Evolution of Stellar Dynamos}",
      journal = {\apj},
         year = 2011,
        month = dec,
       volume = {743},
       number = {1},
          eid = {48},
        pages = {48},
          doi = {10.1088/0004-637X/743/1/48},
archivePrefix = {arXiv},
       eprint = {1109.4634},
 primaryClass = {astro-ph.SR},
       adsurl = {https://ui.adsabs.harvard.edu/abs/2011ApJ...743...48W}
}

@ARTICLE{2014ApJS..211...24M,
       author = {{McQuillan}, A. and {Mazeh}, T. and {Aigrain}, S.},
        title = "{Rotation Periods of 34,030 Kepler Main-sequence Stars: The Full Autocorrelation Sample}",
      journal = {\apjs},
         year = 2014,
        month = apr,
       volume = {211},
       number = {2},
          eid = {24},
        pages = {24},
          doi = {10.1088/0067-0049/211/2/24},
archivePrefix = {arXiv},
       eprint = {1402.5694},
 primaryClass = {astro-ph.SR},
       adsurl = {https://ui.adsabs.harvard.edu/abs/2014ApJS..211...24M}
}

@ARTICLE{2019ApJS..244...21S,
       author = {{Santos}, A.~R.~G. and {Garc{\'\i}a}, R.~A. and {Mathur}, S. and {Bugnet}, L. and {van Saders}, J.~L. and {Metcalfe}, T.~S. and {Simonian}, G.~V.~A. and {Pinsonneault}, M.~H.},
        title = "{Surface Rotation and Photometric Activity for Kepler Targets. I. M and K Main-sequence Stars}",
      journal = {\apjs},
         year = 2019,
        month = sep,
       volume = {244},
       number = {1},
          eid = {21},
        pages = {21},
          doi = {10.3847/1538-4365/ab3b56},
archivePrefix = {arXiv},
       eprint = {1908.05222},
 primaryClass = {astro-ph.SR},
       adsurl = {https://ui.adsabs.harvard.edu/abs/2019ApJS..244...21S}
}

@ARTICLE{2020A&A...635A..43R,
       author = {{Reinhold}, Timo and {Hekker}, Saskia},
        title = "{Stellar rotation periods from K2 Campaigns 0-18. Evidence for rotation period bimodality and simultaneous variability decrease}",
      journal = {\aap},
         year = 2020,
        month = mar,
       volume = {635},
          eid = {A43},
        pages = {A43},
          doi = {10.1051/0004-6361/201936887},
archivePrefix = {arXiv},
       eprint = {2001.08214},
 primaryClass = {astro-ph.SR},
       adsurl = {https://ui.adsabs.harvard.edu/abs/2020A&A...635A..43R}
}

@ARTICLE{2025ApJS..276...57G,
       author = {{Gao}, Xinyi and {Chen}, Xiaodian and {Wang}, Shu and {Liu}, Jifeng},
        title = "{Classification of Periodic Variable Stars from TESS}",
      journal = {\apjs},
         year = 2025,
        month = feb,
       volume = {276},
       number = {2},
          eid = {57},
        pages = {57},
          doi = {10.3847/1538-4365/ad9dd6},
archivePrefix = {arXiv},
       eprint = {2412.06175},
 primaryClass = {astro-ph.SR},
       adsurl = {https://ui.adsabs.harvard.edu/abs/2025ApJS..276...57G}
}

@ARTICLE{2025ApJS..279....5Z,
       author = {{Zhang}, Meng and {Xiang}, Maosheng and {Ting}, Yuan-Sen and {Amarsi}, Anish Mayur and {Zhang}, Hua-Wei and {Shi}, Jianrong and {Yuan}, Haibo and {Li}, Haining and {Wang}, Jiahui and {Wu}, Yaqian and et al.},
        title = "{Homogeneous Stellar Atmospheric Parameters and 22 Elemental Abundances for FGK Stars Derived from LAMOST Low-resolution Spectra with DD-PAYNE}",
      journal = {\apjs},
         year = 2025,
        month = jul,
       volume = {279},
       number = {1},
          eid = {5},
        pages = {5},
          doi = {10.3847/1538-4365/add016},
archivePrefix = {arXiv},
       eprint = {2506.02763},
 primaryClass = {astro-ph.SR},
       adsurl = {https://ui.adsabs.harvard.edu/abs/2025ApJS..279....5Z}
}

@ARTICLE{2013A&A...553A...6H,
       author = {{Husser}, T.-O. and {Wende-von Berg}, S. and {Dreizler}, S. and {Homeier}, D. and {Reiners}, A. and {Barman}, T. and {Hauschildt}, P.~H.},
        title = "{A new extensive library of PHOENIX stellar atmospheres and synthetic spectra}",
      journal = {\aap},
         year = 2013,
        month = may,
       volume = {553},
          eid = {A6},
        pages = {A6},
          doi = {10.1051/0004-6361/201219058},
archivePrefix = {arXiv},
       eprint = {1303.5632},
 primaryClass = {astro-ph.SR},
       adsurl = {https://ui.adsabs.harvard.edu/abs/2013A&A...553A...6H}
}

@ARTICLE{2023MNRAS.519.5271C,
       author = {{Christy}, C.~T. and {Jayasinghe}, T. and {Stanek}, K.~Z. and {Kochanek}, C.~S. and {Thompson}, T.~A. and {Shappee}, B.~J. and {Holoien}, T.~W.-S. and {Prieto}, J.~L. and {Dong}, Subo and {Giles}, W.},
        title = "{The ASAS-SN catalogue of variable stars X: discovery of 116 000 new variable stars using G-band photometry}",
      journal = {\mnras},
         year = 2023,
        month = mar,
       volume = {519},
       number = {4},
        pages = {5271-5287},
          doi = {10.1093/mnras/stac3801},
archivePrefix = {arXiv},
       eprint = {2205.02239},
 primaryClass = {astro-ph.SR},
       adsurl = {https://ui.adsabs.harvard.edu/abs/2023MNRAS.519.5271C}
}

@ARTICLE{2022ApJS..258...16P,
       author = {{Pr{\v{s}}a}, Andrej and {Kochoska}, Angela and {Conroy}, Kyle E. and {Eisner}, Nora and {Hey}, Daniel R. and {IJspeert}, Luc and {Kruse}, Ethan and {Fleming}, Scott W. and {Johnston}, Cole and {Kristiansen}, Martti H. and et al.},
        title = "{TESS Eclipsing Binary Stars. I. Short-cadence Observations of 4584 Eclipsing Binaries in Sectors 1-26}",
      journal = {\apjs},
         year = 2022,
        month = jan,
       volume = {258},
       number = {1},
          eid = {16},
        pages = {16},
          doi = {10.3847/1538-4365/ac324a},
archivePrefix = {arXiv},
       eprint = {2110.13382},
 primaryClass = {astro-ph.SR},
       adsurl = {https://ui.adsabs.harvard.edu/abs/2022ApJS..258...16P}
}

@ARTICLE{2014ApJS..213....9D,
       author = {{Drake}, A.~J. and {Graham}, M.~J. and {Djorgovski}, S.~G. and {Catelan}, M. and {Mahabal}, A.~A. and {Torrealba}, G. and {Garc{\'\i}a-{\'A}lvarez}, D. and {Donalek}, C. and {Prieto}, J.~L. and {Williams}, R. and {Larson}, S. and {Christen sen}, E. and {Belokurov}, V. and {Koposov}, S.~E. and {Beshore}, E. and {Boattini}, A. and {Gibbs}, A. and {Hill}, R. and {Kowalski}, R. and {Johnson}, J. and {Shelly}, F.},
        title = "{The Catalina Surveys Periodic Variable Star Catalog}",
      journal = {\apjs},
         year = 2014,
        month = jul,
       volume = {213},
       number = {1},
          eid = {9},
        pages = {9},
          doi = {10.1088/0067-0049/213/1/9},
archivePrefix = {arXiv},
       eprint = {1405.4290},
 primaryClass = {astro-ph.SR},
       adsurl = {https://ui.adsabs.harvard.edu/abs/2014ApJS..213....9D}
}

@ARTICLE{2022RNAAS...6...96H,
       author = {{Howard}, Erin L. and {Davenport}, James R.~A. and {Covey}, Kevin R.},
        title = "{370 New Eclipsing Binary Candidates from TESS Sectors 1-26}",
      journal = {RNAAS},
         year = 2022,
        month = may,
       volume = {6},
       number = {5},
          eid = {96},
        pages = {96},
          doi = {10.3847/2515-5172/ac6e42},
archivePrefix = {arXiv},
       eprint = {2205.14184},
 primaryClass = {astro-ph.SR},
       adsurl = {https://ui.adsabs.harvard.edu/abs/2022RNAAS...6...96H}
}

@ARTICLE{2017ARep...61...80S,
       author = {{Samus}, N.~N. and {Kazarovets}, E.~V. and {Durlevich}, O.~V. and {Kireeva}, N.~N. and {Pastukhova}, E.~N.},
        title = "{General catalogue of variable stars: Version GCVS 5.1}",
      journal = {AR},
         year = 2017,
        month = jan,
       volume = {61},
       number = {1},
        pages = {80-88},
          doi = {10.1134/S1063772917010085},
       adsurl = {https://ui.adsabs.harvard.edu/abs/2017ARep...61...80S}
}

@ARTICLE{2020ApJS..249...18C,
       author = {{Chen}, Xiaodian and {Wang}, Shu and {Deng}, Licai and {de Grijs}, Richard and {Yang}, Ming and {Tian}, Hao},
        title = "{The Zwicky Transient Facility Catalog of Periodic Variable Stars}",
      journal = {\apjs},
         year = 2020,
        month = jul,
       volume = {249},
       number = {1},
          eid = {18},
        pages = {18},
          doi = {10.3847/1538-4365/ab9cae},
archivePrefix = {arXiv},
       eprint = {2005.08662},
 primaryClass = {astro-ph.SR},
       adsurl = {https://ui.adsabs.harvard.edu/abs/2020ApJS..249...18C}
}

@ARTICLE{2018ApJS..237...28C,
       author = {{Chen}, Xiaodian and {Wang}, Shu and {Deng}, Licai and {de Grijs}, Richard and {Yang}, Ming},
        title = "{Wide-field Infrared Survey Explorer (WISE) Catalog of Periodic Variable Stars}",
      journal = {\apjs},
         year = 2018,
        month = aug,
       volume = {237},
       number = {2},
          eid = {28},
        pages = {28},
          doi = {10.3847/1538-4365/aad32b},
archivePrefix = {arXiv},
       eprint = {1807.03448},
 primaryClass = {astro-ph.SR},
       adsurl = {https://ui.adsabs.harvard.edu/abs/2018ApJS..237...28C}
}

@ARTICLE{2023A&A...674A..14R,
       author = {{Rimoldini}, Lorenzo and {Holl}, Berry and {Gavras}, Panagiotis and {Audard}, Marc and {De Ridder}, Joris and {Mowlavi}, Nami and {Nienartowicz}, Krzysztof and {Jevardat de Fombelle}, Gr{\'e}gory and {Lecoeur-Ta{\"\i}bi}, Isabelle and {Karbevska}, Lea and {Evans}, Dafydd W. and {{\'A}brah{\'a}m}, P{\'e}ter and {Carnerero}, Maria I. and {Clementini}, Gisella and {Distefano}, Elisa and {Garofalo}, Alessia and {Garc{\'\i}a-Lario}, Pedro and {Gomel}, Roy and {Klioner}, Sergei A. and {Kruszy{\'n}ska}, Katarzyna and {Lanzafame}, Alessandro C. and {Lebzelter}, Thomas and {Marton}, G{\'a}bor and {Mazeh}, Tsevi and {Molinaro}, Roberto and {Panahi}, Aviad and {Raiteri}, Claudia M. and {Ripepi}, Vincenzo and {Szabados}, L{\'a}szl{\'o} and {Teyssier}, David and {Trabucchi}, Michele and {Wyrzykowski}, {\L}ukasz and {Zucker}, Shay and {Eyer}, Laurent},
        title = "{Gaia Data Release 3. All-sky classification of 12.4 million variable sources into 25 classes}",
      journal = {\aap},
         year = 2023,
        month = jun,
       volume = {674},
          eid = {A14},
        pages = {A14},
          doi = {10.1051/0004-6361/202245591},
archivePrefix = {arXiv},
       eprint = {2211.17238},
 primaryClass = {astro-ph.GA},
       adsurl = {https://ui.adsabs.harvard.edu/abs/2023A&A...674A..14R}
}

@ARTICLE{2023A&A...674A..34G,
       author = {{Gaia Collaboration} and {Arenou}, F. and {Babusiaux}, C. and {Barstow}, M.~A. and {Faigler}, S. and {Jorissen}, A. and {Kervella}, P. and {Mazeh}, T. and {Mowlavi}, N. and {Panuzzo}, P. and et al.},
        title = "{Gaia Data Release 3. Stellar multiplicity, a teaser for the hidden treasure}",
      journal = {\aap},
         year = 2023,
        month = jun,
       volume = {674},
          eid = {A34},
        pages = {A34},
          doi = {10.1051/0004-6361/202243782},
archivePrefix = {arXiv},
       eprint = {2206.05595},
 primaryClass = {astro-ph.SR},
       adsurl = {https://ui.adsabs.harvard.edu/abs/2023A&A...674A..34G}
}

@ARTICLE{2002ARA&A..40..217G,
       author = {{G{\"u}del}, Manuel},
        title = "{Stellar Radio Astronomy: Probing Stellar Atmospheres from Protostars to Giants}",
      journal = {\araa},
         year = 2002,
        month = jan,
       volume = {40},
        pages = {217-261},
          doi = {10.1146/annurev.astro.40.060401.093806},
archivePrefix = {arXiv},
       eprint = {astro-ph/0206436},
 primaryClass = {astro-ph},
       adsurl = {https://ui.adsabs.harvard.edu/abs/2002ARA&A..40..217G}
}

@ARTICLE{2008ApJ...687.1264M,
       author = {{Mamajek}, Eric E. and {Hillenbrand}, Lynne A.},
        title = "{Improved Age Estimation for Solar-Type Dwarfs Using Activity-Rotation Diagnostics}",
      journal = {\apj},
         year = 2008,
        month = nov,
       volume = {687},
       number = {2},
        pages = {1264-1293},
          doi = {10.1086/591785},
archivePrefix = {arXiv},
       eprint = {0807.1686},
 primaryClass = {astro-ph},
       adsurl = {https://ui.adsabs.harvard.edu/abs/2008ApJ...687.1264M}
}

@ARTICLE{2020ApJ...902..114W,
       author = {{Wang}, Song and {Bai}, Yu and {He}, Lin and {Liu}, Jifeng},
        title = "{Stellar X-Ray Activity Across the Hertzsprung-Russell Diagram. I. Catalogs}",
      journal = {\apj},
         year = 2020,
        month = oct,
       volume = {902},
       number = {2},
          eid = {114},
        pages = {114},
          doi = {10.3847/1538-4357/abb66d},
archivePrefix = {arXiv},
       eprint = {2009.02648},
 primaryClass = {astro-ph.SR},
       adsurl = {https://ui.adsabs.harvard.edu/abs/2020ApJ...902..114W}
}

@ARTICLE{2003A&A...397..147P,
       author = {{Pizzolato}, N. and {Maggio}, A. and {Micela}, G. and {Sciortino}, S. and {Ventura}, P.},
        title = "{The stellar activity-rotation relationship revisited: Dependence of saturated and non-saturated X-ray emission regimes on stellar mass for late-type dwarfs}",
      journal = {\aap},
         year = 2003,
        month = jan,
       volume = {397},
        pages = {147-157},
          doi = {10.1051/0004-6361:20021560},
       adsurl = {https://ui.adsabs.harvard.edu/abs/2003A&A...397..147P}
}

@ARTICLE{2016Natur.535..526W,
       author = {{Wright}, Nicholas J. and {Drake}, Jeremy J.},
        title = "{Solar-type dynamo behaviour in fully convective stars without a tachocline}",
      journal = {\nat},
         year = 2016,
        month = jul,
       volume = {535},
       number = {7613},
        pages = {526-528},
          doi = {10.1038/nature18638},
archivePrefix = {arXiv},
       eprint = {1607.07870},
 primaryClass = {astro-ph.SR},
       adsurl = {https://ui.adsabs.harvard.edu/abs/2016Natur.535..526W}
}

@ARTICLE{2018A&A...618A..48M,
       author = {{Mittag}, M. and {Schmitt}, J.~H.~M.~M. and {Schr{\"o}der}, K.-P.},
        title = "{Revisiting the connection between magnetic activity, rotation period, and convective turnover time for main-sequence stars}",
      journal = {\aap},
         year = 2018,
        month = oct,
       volume = {618},
          eid = {A48},
        pages = {A48},
          doi = {10.1051/0004-6361/201833498},
archivePrefix = {arXiv},
       eprint = {1807.05825},
 primaryClass = {astro-ph.SR},
       adsurl = {https://ui.adsabs.harvard.edu/abs/2018A&A...618A..48M}
}

@ARTICLE{2026ApJ..1000..208H,
       author = {{Han}, Henggeng and {Wang}, Song and {Yang}, Huiqin and {Li}, Xue and {Zheng}, Chuanjie and {Li}, Xiangyu and {Wang}, Cunshi and {Liu}, Jifeng},
        title = "{Varying Core─Envelope Coupling Efficiency Identified from the Stellar Rotation─Activity Relation}",
      journal = {\apj},
         year = 2026,
        month = apr,
       volume = {1000},
       number = {2},
          eid = {208},
        pages = {208},
          doi = {10.3847/1538-4357/ae4aa8},
archivePrefix = {arXiv},
       eprint = {2512.11376},
 primaryClass = {astro-ph.SR},
       adsurl = {https://ui.adsabs.harvard.edu/abs/2026ApJ..1000..208H}
}

@ARTICLE{2025A&A...699A.251Y,
       author = {{Yang}, Huiqin and {Liu}, Jifeng and {Soria}, Roberto and {Spada}, Federico and {Wang}, Song and {Fang}, Xiangsong and {Li}, Xue},
        title = "{Four ages of rotating stars in the rotation─activity relationship and gyrochronology}",
      journal = {\aap},
         year = 2025,
        month = jul,
       volume = {699},
          eid = {A251},
        pages = {A251},
          doi = {10.1051/0004-6361/202554379},
archivePrefix = {arXiv},
       eprint = {2506.07383},
 primaryClass = {astro-ph.SR},
       adsurl = {https://ui.adsabs.harvard.edu/abs/2025A&A...699A.251Y}
}
\bibliographystyle{aasjournal}

\end{document}